\documentclass[10pt,aps,superscriptaddress,twocolumn,floatfix]{revtex4-1}

\usepackage{color}
\usepackage{graphicx}
\usepackage{amsmath,amssymb,array,mathtools}
\usepackage{leftidx}
\usepackage{subfigure}
\usepackage[normalem]{ulem}
\usepackage{appendix}
\usepackage{array}
\usepackage{booktabs}
\usepackage{multirow}

\newcommand{\be} {\begin{equation}}
\newcommand{\ee} {\end{equation}}
\newcommand{\bea} {\begin{eqnarray}}
\newcommand{\eea} {\end{eqnarray}}
\newcommand{\vctr}[1]{\mathbf{#1}}

\newcommand{\bilu} {\beta^{I}_{\mu}}
\newcommand{\bild} {\beta^{I}_{\nu}}
\newcommand{\qq} {q^{I}_{\mu\nu}}
 
\newcommand{\bra}[1] { \langle #1 | }
\newcommand{\ket}[1] { | #1 \rangle }
\newcommand{\braket}[2] { \langle #1 | #2 \rangle }
\newcommand{\Bra}[1] { \Bigl\langle #1 \Bigr| }
\newcommand{\Ket}[1] { \Bigl| #1 \Bigr\rangle }

\renewcommand{\vr} {{\bf r}}

\newcommand{\Proj} {\hat{P}^I_{m_2 m_1}}
\newcommand{\Projj} {\hat{P}^I_{m_1 m_2}}
\newcommand{\Prj} {\hat{P}}

\newcommand{\psis}{\psi_{i\sigma}}
\newcommand{\psjs}{\psi_{j\sigma}}
\newcommand{\Dpsis}{\widetilde{\Delta}^\mu\psi_{i\sigma}}
\newcommand{\DpsisE}{\widetilde{\Delta}^\alpha_E \psi_{i\sigma}}
\newcommand{\Dpsismet}{\Delta^\mu\psi_{i\sigma}}
\newcommand{\Dn}{\widetilde{\Delta}^\mu n^{I\sigma}_{m_1m_2}}
\newcommand{\DnE}{\widetilde{\Delta}^\alpha_E n^{I\sigma}_{m_1m_2}}
\newcommand{\Dnmet}{\Delta^\mu n^{I\sigma}_{m_1m_2}}
\newcommand{\sumi}{\sum_{i}}
\newcommand{\sumis}{\sum_{i\sigma}}
\newcommand{\sumj}{\sum_{j}}

\newcommand{\sumIsmm}{\sum_{I\sigma m_1 m_2}}
\newcommand{\eis}{\varepsilon_{i\sigma}}
\newcommand{\ejs}{\varepsilon_{j\sigma}}

\newcommand{\nn}{\nonumber}
\newcommand{\f}{\varphi} 
\newcommand{\de}{\partial}
\newcommand{\vtilde}{V_U^{\sigma}}
\newcommand{\vtildeImm}{V_{U, m_1 m_2}^{I \sigma}}

\newcommand{\thetisjs}{\theta_{i\sigma,j\sigma}}
\newcommand{\thetjsis}{\theta_{j\sigma,i\sigma}}
\newcommand{\thetFis}{\tilde{\theta}_{F,i\sigma}}
\newcommand{\thetFjs}{\tilde{\theta}_{F,j\sigma}}
\newcommand{\editor}[2]{%
  \expandafter\newcommand\csname #1note\endcsname[1]{%
    \textcolor{#2}{(\textbf{#1:} ##1)}}%
  \expandafter\newcommand\csname #1\endcsname[1]{%
    \textcolor{#2}{##1}}%
  \expandafter\newcommand\csname #1cancel\endcsname[1]{%
    \textcolor{#2}{\sout{##1}}}%
  \expandafter\newcommand\csname #1change\endcsname[2]{%
    \textcolor{#2}{\sout{##1} ##2}}%
  \newenvironment{#1text}{\color{#2}}{\color{black}}
}

\definecolor{red}{rgb}{1.00,0.00,0.00}
\definecolor{green}{rgb}{0.00,1.00,0.00}
\definecolor{blue}{rgb}{0.00,0.00,1.00} 
\definecolor{orange}{rgb}{1.00,0.50,0.00}
\definecolor{ORANGE}{rgb}{1.00,0.50,0.00}
\definecolor{magenta}{rgb}{1.00,0.00,1.00}
\definecolor{cyan}{rgb}{0.00,1.00,1.00}
\definecolor{brown}{rgb}{0.4,0.2,0.00}
\definecolor{deepsky}{rgb}{0.0,0.75,1.00}
\definecolor{gray}{rgb}{0.50,0.50,0.50}
\definecolor{navy}{rgb}{0.137,0.137,0.557}
\definecolor{burntorange}{rgb}{0.8, 0.33, 0.0}
\definecolor{asparagus}{rgb}{0.53, 0.66, 0.42}
\definecolor{emerald}{rgb}{0, 0.605, 0.465}
\definecolor{amethyst}{rgb}{0.6, 0.4, 0.8}
\definecolor{purple}{rgb}{0.502, 0, 0.502}
\definecolor{lava}{rgb}{0.81, 0.06, 0.13}
\definecolor{tk}{rgb}{0.5, 0.00, 0.8}
\definecolor{beaver}{rgb}{0.62, 0.51, 0.44}
\definecolor{yellow}{rgb}{0.7, 0.7, 0.0}
\definecolor{seagreen}{rgb}{0.18,0.55,0.34}
\definecolor{oxford}{rgb}{0.0,0.129,0.278}
\definecolor{lightgray}{rgb}{.9,.9,.9}
\definecolor{lightblue}{rgb}{0.0,0.0,0.8}
\definecolor{fuchsia}{rgb}{1.0,0.47,1.0}

\editor{AF}{red}
\editor{MC}{blue}
\editor{IT}{burntorange}
\editor{SdG}{orange}
\editor{NM}{seagreen}

\begin{document}

\title{Hubbard-corrected density functional perturbation theory with ultrasoft pseudopotentials}

\author{A. Floris}\email[e-mail:]{
afloris@lincoln.ac.uk} 
\affiliation{School of Chemistry, University of Lincoln, Brayford Pool, Lincoln LN6 7TS, United Kingdom}

\author{I. Timrov}
\affiliation{Theory and Simulation of Materials (THEOS) and National Centre for Computational Design and Discovery of Novel Materials (MARVEL), \'Ecole Polytechnique F\'ed\'erale de Lausanne (EPFL), CH-1015 Lausanne, Switzerland}

\author{B. Himmetoglu}
\affiliation{Department of Chemical Engineering and Materials Science,
University of Minnesota, 421 Washington Ave SE, Minneapolis, MN 55455, USA} 

\author{N. Marzari}
\affiliation{Theory and Simulation of Materials (THEOS) and National Centre for Computational Design and Discovery of Novel Materials (MARVEL), \'Ecole Polytechnique F\'ed\'erale de Lausanne (EPFL), CH-1015 Lausanne, Switzerland}

\author{S. de Gironcoli}
\affiliation{Scuola Internazionale Superiore di Studi Avanzati (SISSA), Via Bonomea 265, 34136 Trieste, Italy}
\affiliation{CRS Democritos, CNR-IOM Democritos, Via Bonomea 265, 34136 Trieste, Italy}

\author{M. Cococcioni}
\email[e-mail:]{ matteo.cococcioni@unipv.it}

\affiliation{Department of Physics, University of Pavia, Via A. Bassi 6, I-27100 Pavia, Italy}

\date{\today}

\begin{abstract}
We present in full detail a newly developed formalism 
enabling density functional 
perturbation theory (DFPT) calculations from a DFT+$U$ ground state. 
The implementation 
includes ultrasoft pseudopotentials and is valid for both insulating and metallic systems. 
It aims at fully 
exploiting the versatility of DFPT combined with the low-cost DFT+$U$ functional. 
This allows us to avoid computationally 
intensive frozen-phonon calculations when DFT+$U$ is used to eliminate the residual electronic self-interaction from approximate functionals and to capture the localization of   valence electrons e.g. on $d$ or $f$ states. In this way, the effects of electronic localization (possibly due to correlations) are consistently taken into account
in the calculation of specific phonon modes, Born effective charges, dielectric
tensors and in quantities requiring well converged sums over many phonon 
frequencies, as phonon density of states and free energies. 
The new computational tool is applied to two representative systems, namely CoO, a prototypical transition metal monoxide and LiCoO$_2$, a material employed for the cathode of Li-ion batteries. 
The results show the effectiveness of our formalism to capture in a quantitatively reliable way the vibrational properties of systems with localized valence electrons. 
\end{abstract}

\pacs{}

\maketitle

\section{Introduction }
\label{intro}

Systems characterized by a pronounced localization of valence electrons (typically $d$ or $f$) still represent a significant challenge for current implementations of density functional theory (DFT)~\cite{hohenberg64, kohn65}, and most of the available approximations to the exact exchange-correlation (xc) functional fail quite spectacularly in describing their physical properties. Several methodologies have been developed to avoid (or at least alleviate) the over-delocalization of valence electrons and its catastrophic consequences on the quality of the predicted ground state. 
Some of these techniques aim at the direct elimination of the residual self-interaction, 
either through a self-interaction corrected (SIC) functional 
~\cite{Perdew:1981,heaton83,dabo10}, or through the addition of various amounts of  Fock exact exchange, as in hybrid functionals~\cite{blyp88,becke93,perdew96,Heyd:2003}.
Some others (often outside the DFT theoretical framework) target instead a better representation of electronic correlations, e.g. through
the mapping of the electronic problem onto a suitable local model solved with many-body techniques, as in dynamical mean-field theory (DMFT)~\cite{metzner89,georges92,jarrell92,georges96}; or using an extended formulation in case delocalization is the result of degeneracy, as in  ensemble DFT~\cite{theophilou79,Gross:1988,Gross:1988a}; or employing a more structured variable as in reduced density-matrix functional theory (RDMFT)~\cite{Gilbert:1975, Lathiot:2010, Sharma:2013}. 
Almost invariably, however, all these approaches have  computational costs  significantly higher than those of standard, approximate DFT functionals.

Among the corrective schemes
defined to alleviate  the consequences of the residual self-interaction, the DFT+$U$ method~\cite{anisimov91,anisimov96,anisimov97,dudarev98} is one of the most widely used. Its popularity is mostly due to a very simple formulation and a low computational cost, two factors that make its implementation straightforward and offer the unique possibility to study systems whose size and complexity would be prohibitive with more sophisticated methods. A very distinctive  advantage brought about by the DFT+$U$ formulation is the possibility to analytically derive and  easily implement the derivatives of the total energy (forces, stresses, dynamical matrices, etc.) and  calculate them accurately and efficiently. These quantities are necessary to identify the equilibrium structure of materials, their elastic properties, their vibrational spectrum and to account for finite-temperature effects on properties of interest. 

In Ref.~\cite{floris11}, a formalism was introduced to compute the vibrational properties of strongly correlated systems, based on the extension of density functional perturbation theory (DFPT)~\cite{baroni87,giannozzi91,gonze95,gonze95_1,baroni01} to the DFT+$U$ functional. The resulting approach was acronymed DFPT+$U$. MnO and NiO phonon dispersions calculated with DFPT+$U$~\cite{floris11} demonstrated that a better representation of electronic localization can improve significantly the agreement with available experimental data, with respect to uncorrected DFT functionals as, e.g., those based on the generalized gradient approximation (GGA). In fact, not only the values of vibrational frequencies were improved, but also the width of the splitting between the longitudinal and transverse optical phonon frequencies, due to the peculiar antiferromagnetic order that characterizes  these systems~\cite{massidda}. DFPT+$U$ has also been used to compute phonon-related
properties of various Earth's minerals, overall improving their vibrational spectra~\cite{shukla15,blanchard14,blanchard15,shukla16,shukla16g}. Improvements in Raman spectra of strongly correlated materials have been discussed also in a recent (independent) implementation of DFPT+$U$~\cite{Miwa:2018}.

In this work we extend the DFPT+$U$ formalism to ultrasoft (US) pseudopotentials (PPs)~\cite{vanderbilt90}, 
discussing in full detail the additional contributions needed with respect to the norm-conserving (NC) PPs formalism~\cite{floris11}. 
In addition, we generalize the DFPT+$U$ formulation to metallic ground states, characterized by a finite smearing in the Fermi-Dirac distribution function. This generalization is useful to study metallic materials~\cite{Note_tech} where some properties depend critically on the localization of valence $d$ electrons as, for example, the magnetic properties and the phase stability of transition-metal compounds (e.g., Heusler alloys \cite{himmetoglu12} or oxides) or for systems whose degenerate ground state results in partially filled Kohn-Sham (KS) orbitals, as FeO~\cite{himmetoglu14}. Moreover, the metallic formalism is crucial to explore possible effects of the Hubbard $U$ on the electron-phonon coupling.
The extension of DFPT+$U$ to US PPs and metallic ground states presented in this work is based on the one introduced in Ref.~\cite{dalcorso01} for DFPT,
whose formalism is adopted here. 
An analogous independent adaptation of DFPT+$U$ to US PPs was recently derived in Ref.~\cite{Miwa:2018} which, however, does not develop the metallic case. In addition, while Ref.~\cite{Miwa:2018} uses the formulation of DFT+$U$ introduced in Ref.~\cite{liech95} that depends explicitly on the on-site Hubbard interaction $U$ and the exchange coupling $J$,  
the present work is based on the simplified  formulation of DFT+$U$ introduced in Ref.~\cite{dudarev98} that depends only on the ``effective $U$" (broadly corresponding to $U-J$). 
More importantly, while Ref.~\cite{Miwa:2018} is based on the projection of KS states on the
projector functions (see Eq.~(6) of Ref.~\cite{vanderbilt90})
in the augmentation spheres 
this work uses projections on atomic orbitals, thus achieving a straight generalization of the implementation for NC PPs~\cite{floris11}.

The DFPT+$U$ formulation including US PPs combines the efficiency of both DFPT and US PPs, making the calculation of DFT+$U$  linear-response quantities straightforward and accurate. In phonon calculations (the main focus of this paper), DFPT+$U$ is used to capture the effects of electronic localization on vibrational frequencies, modes, Born effective charges, and dielectric tensors, and will provide full access to all quantities requiring well converged sums over the entire vibrational spectrum, as phonon density of states, free energies, electron-phonon coupling, and thermal transport.

We demonstrate the effectiveness of  DFPT+$U$ by computing the phonon dispersions of two Co compounds, namely CoO (one of the prototypical transition-metal monoxides), and LiCoO$_2$ (a layered oxide used as cathode of Li-ion batteries~\cite{Ozawa:1994, Takahashi:2008}). These quite different systems, where Co appears in the 2+ and 3+ oxidation states, respectively, will illustrate the importance of the Hubbard correction to capture the localization of $3d$ electrons, and its effects on ground-state and  vibrational properties. In LiCoO$_2$ the Hubbard correction improves sensibly the agreement with experimental data for both the equilibrium crystal structure and the band gap and imposes a nonuniform blue-shift of the highest part of the vibrational spectrum. Overall, this improves the quantitative agreement with the measured frequencies at the $\Gamma$ point.
The effect of the Hubbard correction on the properties of CoO is far more radical. Here the Hubbard $U$ eliminates the fictitious metallic ground state achieved with the GGA functional, stabilizing an insulating ground state with a finite band gap. This not only  improves dramatically the equilibrium crystal structure compared to experiments, but resolves all the dynamical instabilities obtained from GGA and accounts for the splittings between the TO modes (due to the antiferromagnetic order), which is qualitatively consistent with those previously obtained for MnO and NiO~\cite{floris11}.

The 
paper is organized as follows. In Sec.~\ref{single} we summarize the DFT+$U$ formalism in the US PPs context; in Sec.~\ref{forces} we present the first derivatives of the total energy, with particular emphasis to the Hubbard forces; in Sec.~\ref{linear} we discuss the DFPT+$U$ formalism, namely the DFPT extension to the Hubbard functional (whose adaptation to metals is developed in the  
Appendix);
in Sec.~\ref{sec:technical_details} we present the technical details of our calculations; in
Sec. \ref{validation}  our formalism is employed to study the vibrational properties of CoO and LiCoO$_2$; finally, in Sec.~\ref{concl} we give our conclusive remarks.
Benchmarks of our implementation are presented in the Supplemental Material (SM)~\cite{SupplementalMaterial} (see also references \cite{Dalcorso:2014, Dalcorso:2000, Davey:1925, phonopy}).  
We use Hartree atomic units throughout the paper.

\section{DFT+$U$}
\label{single}

In DFT+$U$, the DFT total energy  $E_\mathrm{DFT}$ is augmented by a corrective term, namely the Hubbard energy $E_U$:
\be
E_{\mathrm{DFT}+U} = E_\mathrm{DFT} + E_U \,. 
\label{edftu}
\ee
The $E_U$ expression is shaped on the Hubbard model. 
We adopt here the simplest rotationally invariant formulation introduced in Ref.~\cite{dudarev98} and constructed within the so-called fully-localized limit (FLL) of the double-counting term. Within this approximation the corrective energy $E_U$ reads: 
\be
\label{ehub}
E_U = \frac{1}{2} \sum_{I\sigma m_1 m_2} U^I (\delta_{m_1 m_2} -n^{I\sigma}_{m_1 m_2}) \, n^{I\sigma}_{m_2 m_1} \,,  
\ee
where $I$ is the atomic site index, $\sigma$ is the spin index, $m_1$ and $m_2$ are magnetic quantum numbers associated with a specific angular momentum; 
$U^I$ is the effective Hubbard parameter of the $I$-th atom. 
Other formulations of the $E_U$ functional 
are also popular in literature, including different recipes for the double-counting term \cite{czyzyk94,chioncel03} and various formulations of the Hubbard part, most notably the one introduced in Ref.~\cite{liech95}, featuring a 
Hatree-Fock-like interaction term with screened Coulomb and exchange couplings. 
The results obtained in this work can be easily extended to these 
formulations, once the various terms of the derivatives are recast accordingly.

In Eq.~\eqref{ehub}, $n^{I\sigma}_{m_1 m_2}$ are the occupation matrices (real and symmetric) of the atomic orbitals $\f^I_{m}(\mathbf{r})$ that form the localized basis set used in the Hubbard corrective Hamiltonian. Their generalized US PPs expression imposes reviewing the most important features of the US pseudization~\cite{vanderbilt90} (in the following we will refer to pseudo-wave functions simply as ``wave functions''). 

In the US formalism, the wave functions are obtained from their all-electron counterparts, by smoothing the fastest oscillations that they exhibit in the atomic core region. Crucially, during this procedure the normalization of the wave functions is lost. Quantities like the electronic charge density or the scalar products between wave functions need to be corrected to compensate for the missing parts. Their ``augmentation" is realized through suitably defined projectors onto the atomic core regions, where the smoothing of the wave functions occurs.
The electronic charge density is thus computed as follows:
\begin{eqnarray}
\rho(\vr) & = & \sumis \biggl[ \vert\psis(\vr)\vert^2  \biggl.\nonumber \\
& & \hspace{1cm} \biggl. + \, \sum_{I\mu\nu} Q^I_{\mu \nu}(\vr - \mathbf{R}_I)
\langle\psis\ket{\bilu}\bra{\bild}\psis\rangle \biggr] \nonumber \\
& = & \sumis \bra{\psis} \hat{K}(\vr) \ket{\psis} \,,
\label{rhousk}
\end{eqnarray}
where $\psis(\vr)$ are the KS wave functions labeled by index $i$; $\bilu$ and $\bild$, labeled by an atomic ($I$) and a state (greek letter) index, are localized projector functions that are non-zero only within the augmentation sphere of the atom $I$; 
$Q^I_{\mu \nu}(\vr-{\bf R}_I)$ are the augmentation functions, which contain the difference between the all-electron and pseudo charge densities around the atom at ${\bf R}_I$. The second equality in Eq.~\eqref{rhousk} defines a more compact notation by introducing the ``augmentation" kernel $\hat{K}(\vr)$, which in the full coordinate representation is a function of three spatial coordinates~\cite{dalcorso01}
\begin{eqnarray}
K(\mathbf{r};\mathbf{r}_1, \mathbf{r}_2) & = & \delta(\mathbf{r} - \mathbf{r}_1) \delta(\mathbf{r} - \mathbf{r}_2) +\sum_{I\mu \nu} Q^I_{\mu \nu}(\vr-{\bf R}_I) \nonumber \\
& & \times \, \beta_\mu^I(\mathbf{r}_1 - {\bf R}_I) \, \beta_\nu^{*I}(\mathbf{r}_2 - {\bf R}_I) 
\label{kernel}
\end{eqnarray}
and carries a dependence on the atomic 
coordinates ${\bf R}_I$'s. Scalar products between wave functions in the US PPs scheme are corrected as:
\be
\bra{\psis} \hat{S} \ket{\psjs} = \braket{\psis}{\psjs} + \sum_{I\mu\nu} \qq 
\braket{\psis}{\bilu} \braket{\bild}{\psjs} \,,
\label{scprod}
\ee
where the coefficients $\qq$ are integrals (over the volume of the crystal) of the augmentation functions: $\qq \equiv \int Q^I_{\mu \nu}(\vr) \, d\vr$. The orthonormalization condition between KS states is then generalized as follows:
\be
\bra{\psis} \hat{S} \ket{\psi_{j\sigma'}}=\delta_{ij}\delta_{\sigma\sigma'} \,.
\label{ortho}
\ee
The operator $\hat{S}$, implicitly defined via Eq.~\eqref{scprod}, acts as an ``overlap kernel"
and has the explicit expression~\cite{vanderbilt90}: 
\be
\hat{S} = 1 + \sum_{I\mu\nu}  \qq \ket{\bilu} \bra{\bild} \,.
\label{smatcom}
\ee
The occupation matrices in Eq.~\eqref{ehub} are obtained from the projection of the occupied KS wave functions on the atomic orbitals $\f^I_{m}(\mathbf{r})$ of the DFT+$U$ localized basis set, and, based on the generalization of the scalar products outlined above, take the expression:
\be
n^{I\sigma}_{m_1m_2} = \sumi \bra{\psis} \Proj \ket{\psis} \label{ns} \,,
\ee
where $\Proj$ is the generalized projector on the manifold of localized (atomic) orbitals: 
\be
\Proj = \hat{S} \ket{\f^I_{m_2}}\bra{\f^I_{m_1}} \hat{S} \,.
\label{proj}
\ee
In the presence of a finite overlap between wave functions [Eq.~\eqref{ortho}] the KS equations are generalized as follows:
\be 
\hat{H}^\sigma \ket{\psis} = \eis \hat{S} \ket{\psis} \,,
\label{ks}
\ee
where $\hat{H}^\sigma$ is the Hamiltonian of the system
\be
\hat{H}^\sigma \equiv -\frac{1}{2}\nabla^2 + \hat{V}^{\sigma} \,,
\label{eq:Hamiltonian}
\ee
and $\eis$ are the KS eigenvalues. Following the notation of Ref.~\cite{dalcorso01}, the single-particle total potential in Eq.~\eqref{eq:Hamiltonian}, can be written as:
\be 
\hat{V}^{\sigma} = \hat{V}_\mathrm{KS}^{\sigma} + \hat{V}^{\sigma}_\mathrm{Hub} \,,
\label{pot}
\ee
where 
\bea 
\hat{V}_\mathrm{KS}^{\sigma} = \hat{V}_\mathrm{NL} + \int \hat{K}(\vr) \, V^{\sigma}_\mathrm{eff}(\vr) \, d\mathbf{r} \,,
\label{kspot}
\eea
is the KS potential. This includes the non-local part of the US PP, $\hat{V}_\mathrm{NL}$, and the local 
effective potential $\hat{V}^{\sigma}_\mathrm{eff}$~\cite{vanderbilt90}, which  reads:
\be
V^{\sigma}_\mathrm{eff}(\vr)= V_\mathrm{loc}(\vr) + V^{\sigma}_\mathrm{Hxc}(\vr) \,.
\label{veff}
\ee
Here, $V_\mathrm{loc}(\vr)$ is the local part of the PP, and $V^{\sigma}_\mathrm{Hxc}(\vr) = V_\mathrm{H}(\vr) + V^{\sigma}_\mathrm{xc}(\vr)$ is the sum of Hartree and xc potentials. 
Finally, $\hat{V}^{\sigma}_\mathrm{Hub}$ is the non-local Hubbard potential which reads: 
\be 
\hat{V}^{\sigma}_\mathrm{Hub} = \sum_{I m_1 m_2} \vtildeImm \, \Projj \equiv V^\sigma_U \hat P\,, 
\label{vhub}
\ee
where $\vtildeImm$ is defined as follows
\be
\vtildeImm = U^I \left(\frac{\delta_{m_1 m_2}}{2} - n^{I\sigma}_{m_1 m_2}\right) \,,
\label{vtilde}
\ee
and the projector $\Projj$ was defined in Eq.~\eqref{proj}.
The last equality in Eq.~\eqref{vhub} introduces a notation that will be useful in the following.

In the next section we will see how the above formalism can be used to derive the Hubbard forces,  needed not only for a structural optimization in the framework of DFT+$U$ but also to derive various quantities of the DFPT+$U$ formalism.

\section{Hubbard forces}
\label{forces}

The evaluation of the first-order derivatives of the total energy, i.e. the atomic forces, is the necessary first step for the calculation of all higher order derivatives. 
Starting from the expression of $E_{\mathrm{DFT}+U}$, Eq.~\eqref{edftu}, it can be shown that the force $F_{\lambda}$ acting on the $I$-th atom in the direction $\alpha$ ($\lambda \equiv \{I\alpha\}$) can be computed using the following expression~\cite{Note_ortho,Note_Ewald}:
\begin{eqnarray}
F_{\lambda} & = & - \frac{d E_{\mathrm{DFT}+U}}{d \lambda} \nonumber \\
& = & - \sumis 
\Bra{\psis} \left[ \frac{\de \hat{V}_\mathrm{KS}^{\sigma}}{\de \lambda} 
- \eis \frac{\de \hat{S}}{\de\lambda} \right] \Ket{\psis} - \frac{\de E_U}{\de\lambda} \,.
\label{for}
\end{eqnarray}
The use of US PPs thus introduces an extra term to the forces, namely the
derivative $\sumis 
\Bra{\psis} \eis \frac{\de \hat{S}}{\de\lambda} \Ket{\psis}$, due to the generalized orthonormality condition of KS wave functions, Eq.~\eqref{ortho}. Following Ref.~\cite{dalcorso01},  $\frac{\de }{\de\lambda}$ indicates a {\it bare} derivative that does not involve the response of the KS wave functions $\psis$; $\frac{d}{d\lambda}$ indicates instead a {\it total} derivative which contains also the response of KS wave functions. Bare derivatives, based on the explicit dependence on 
atomic positions $\{{\bf R}^I\}$ only require the knowledge of the unperturbed  KS wave functions.  Total derivatives require instead to evaluate the response of the KS wave functions and need to be re-computed at each DFPT iteration, as shown in Secs.~\ref{linear} and in the Appendix; for this reason they are also called {\it self-consistent-field} (SCF) derivatives. Although defined as {\it total} derivatives of the energy, the calculation of the forces only involves {\it bare} derivatives, as a result of the Hellmann-Feynman theorem and its generalization to US PPs~\cite{Note_ortho}, given in Eq.~\eqref{for}.

The bare derivative of the KS potential in Eq.~\eqref{for} reads~\cite{dalcorso01,Note_force_cancellation}:
\bea 
\frac{\de \hat{V}^{\sigma}_\mathrm{KS}} {\de\lambda} & = & 
\frac{\de \hat{V}_\mathrm{NL}}{\de \lambda} + 
\int \hat{K}(\vr) \, \frac{ \de V_\mathrm{loc}(\vr)}{\de \lambda}  \, d\mathbf{r}  \nn \\ 
& & + \, \int \frac{\de \hat{K}(\vr)}{\de \lambda} \, V^{\sigma}_\mathrm{eff}(\vr) \, d\mathbf{r} \,. 
\label{dvbare}
\eea
The last piece of Eq.~\eqref{for} is the so-called {\it Hubbard force}, and contains the Pulay's terms originating from the shift of the centers of the atomic orbitals. This term was also discussed in Refs.~\cite{himmetoglu14,cococcioni10}. Using Eq.~\eqref{ehub}, it can be expressed in terms of the bare derivative of the occupation  matrices: 
\be
\frac{\de E_U}{\de\lambda}= \sum_{I \sigma m_1 m_2} U^I \left(\frac{\delta_{m_1 m_2}}{2}-n^{I\sigma}_{m_1 m_2}\right)\frac{\de n^{I\sigma}_{m_2 m_1} }{\de \lambda} \,,
\label{hubfor}
\ee
where, based on Eqs.~\eqref{ns} and \eqref{proj}, we have:
\be
\frac{\de n^{I\sigma}_{m_2 m_1} }{\de \lambda} = \sumi   
\Bra{\psis} \frac{\de \Projj}{\de \lambda} \Ket{\psis} \,,
\label{dnbare}
\ee
with
\begin{eqnarray}
\frac{\de \Projj}{\de \lambda} & = & 
\Ket{\frac{ \de(\hat{S}\f^I_{m_1})}{\de \lambda}} \bra{\f^I_{m_2}} \hat{S} \nn \\
& & \hspace{1cm} + \, \hat{S} \ket{\f^I_{m_1}} \Bra{\frac{\de (\hat{S} \f^I_{m_2})} {\de \lambda}} \,.
\label{dproj}
\end{eqnarray}
By virtue of Eqs.~(\ref{vtilde}), (\ref{hubfor}), (\ref{dnbare}), and (\ref{dproj}) the Hubbard force can  
be rewritten as:
\be
 \frac{\de E_U}{\de\lambda} = \sumis 
\Bra{\psis} \vtilde \frac{\de \Prj }{\de \lambda} \Ket{\psis} \,, 
\label{Ehub_bare_deriv}
\ee
where the same short-hand notation of Eq.~\eqref{vhub} is used:
\be
\vtilde \frac{\de \Prj}{\de \lambda} \equiv 
\sum_{I m_1 m_2} \vtildeImm \frac{\de \Projj }{\de \lambda}  \,.
\label{VP_shorthand}
\ee
Equation~\eqref{for} is then recast in the form 
\be
F_{\lambda} = -\sumis 
\Bra{\psis} \left[ \frac{\de \hat{V}_\mathrm{KS}^{\sigma}}{\de \lambda} + 
\vtilde \frac{\de \Prj }{\de \lambda} - \eis \frac{\de \hat{S}}{\de\lambda} \right] \Ket{\psis} \,,
\label{for2}
\ee
which represents the final expression of the force.

\section{DFPT+$U$}
\label{linear}

We now present the calculation of the second order derivatives of the total energy and, in particular, the matrix of force constants. Consistently with the aim of the paper, the discussion is focused on the generalization of the DFPT+$U$ formalism introduced in Ref.~\cite{floris11} to US PPs.
However, it is important to remark that the formalism developed here can be also used, with no extra terms, with projector augmented wave (PAW) PPs~\cite{blochl94}, once US PPs quantities are properly substituted by PAW ones~\cite{Kresse:1999,audouze08}. A validation for PAW PPs is contained in the Supplemental Material~\cite{SupplementalMaterial}, Sec.~\ref{secSM:insulators}. It is important to remark that our implementation is different from most implementations for US and PAW PPs 
(see, e.g., Refs.~\cite{bengone00,Sclauzero:2013,Miwa:2018,Rohrbach:2003}), because it is based on projections on atomic orbitals $\varphi^I$ rather than on projector functions $\beta^I$ (localized inside the augmentation spheres). 


\subsection{Response Hubbard potential and occupation matrix}
\label{sec:pert_Hub_pot}

The matrix of atomic force constants is defined as the second derivative of the total energy with respect to the atomic displacement or, equivalently, as the (negative) first derivative of the atomic forces. From the expression of forces in Eq.~\eqref{for2} it is easy to realize that the calculation of second derivatives implies computing, besides other terms, the derivative (response) of the KS wave functions with respect to atomic displacements $\frac{d\psis(\vr)}{d\mu}$. 
These quantities can be obtained by solving the Sternheimer equation stemming from a first-order expansion of the KS equations~\eqref{ks}:
\begin{eqnarray} 
& & \left[-\frac{1}{2}\nabla^2 + \hat{V}^{\sigma} - \eis \hat{S} \right] \Ket{\frac{d\psis}{d\mu}}  \nn  \\ 
& & \hspace{0.7cm} = - \left[ \frac{d \hat{V}^{\sigma}}{d \mu} - \eis \frac{\de \hat{S}}{\de \mu} - \frac{d \eis}{d \mu} \hat{S} \right]  \ket{\psis} \,.
\label{dks}
\end{eqnarray}
Since the variation of the total potential $\frac{d \hat{V}^{\sigma}}{d \mu}$ depends, by virtue of the Hohenberg-Kohn theorem~\cite{hohenberg64}, on the charge density response (which in turn depends on the wave functions response), Eq.~\eqref{dks} needs to be solved self-consistently in $\frac{d\psis(\mathbf{r})}{d\mu}$. Note that Hubbard corrections enter both $\hat{V}^{\sigma}$ [Eq.~\eqref{pot}] and $\frac{d \hat{V}^{\sigma}}{d \mu}$ and, through the SCF solution of Eq.~\eqref{dks}, affect both $\frac{d\psis(\mathbf{r})}{d\mu}$ and $\frac{d\rho(\vr)}{d\mu}$.

Due to the invariance of the energy 
functional with respect to unitary transformations in the occupied 
manifold, for improved stability the Sternheimer equation~\eqref{dks} is typically solved for the conduction component of $\frac{d\psis(\mathbf{r})}{d\mu}$
\be
\ket{\Dpsis} \equiv \hat{\mathcal{P}}_{c,\sigma} \Ket{\frac{d\psis}{d\mu}},
\label{projc}
\ee
where $\hat{\mathcal{P}}_{c,\sigma}$ is the projector operator on the conduction manifold.
In practice, $\ket{\Dpsis}$ is obtained by applying $\hat{\mathcal{P}}_{c,\sigma}^{\dagger}$ to both  sides of Eq.~\eqref{dks} and solving the equation~\cite{PHnote}:  
\begin{eqnarray} 
& & \left[-\frac{1}{2}\nabla^2 + \hat{V}^{\sigma} 
- \eis \hat{S} \right] \ket{\Dpsis}  \nn  \\ 
& & \hspace{1.3cm} = - \hat{\mathcal{P}}^\dagger_{c,\sigma} \left[ \frac{d \hat{V}^{\sigma}}{d \mu} - \eis \frac{\de \hat{S}}{\de \mu} \right]  \ket{\psis} \,.
\label{dks2}
\end{eqnarray}
$\hat{\mathcal{P}}_{c,\sigma}$ is conveniently computed exploiting the identity 
\be
\hat{\mathcal{P}}_{c,\sigma} = 1 - \hat{\mathcal{P}}_{v,\sigma} = 1 - \sumj^\mathrm{occ} | \psjs \rangle \langle \psjs | \hat{S} \,,
\label{projv}
\ee
where $\hat{\mathcal{P}}_{v,\sigma}$ is the projector operator on the valence manifold. Crucially, the above identity allows to avoid slowly converging sums over conduction states~\cite{Note_LHS_Sternheimer}. 

While within the NC PP formalism {\it only} $\ket{\Dpsis}$ contributes to the total variation of the charge density $\frac{d\rho(\vr)}{d\mu}$ \cite{baroni01}, in the US PP case, due to the generalized orthogonality condition [Eq.~\eqref{ortho}], also the  component of $\frac{d\psis(\mathbf{r})}{d\mu}$  on the valence  manifold  contributes, but only with bare terms~\cite{dalcorso01}, as illustrated below. 

The total variation of the single-particle potential in Eq.~\eqref{dks2} reads:
\be
\frac{d \hat{V}^{\sigma}} {d \mu} = \frac{d \hat{V}^{\sigma}_\mathrm{KS}}{d\mu} + 
\frac{d \hat{V}^{\sigma}_\mathrm{Hub}} {d \mu} \,, 
\label{dvtot}
\ee
where $\frac{d \hat{V}^{\sigma}_\mathrm{KS}} {d\mu}$ is the derivative of the KS potential  present in ``standard'' DFPT implementations~\cite{baroni01,dalcorso01}. $\frac{d \hat{V}^{\sigma}_\mathrm{Hub}} {d \mu}$ is the response of the Hubbard potential: This is the \textit{first} extra term needed in DFPT+$U$. From Eq.~\eqref{vhub} this total derivative is the sum of two contributions:
\be
\frac{d \hat{V}^{\sigma}_\mathrm{Hub}}{d\mu} = \vtilde \frac{\de \Prj}{\de \mu} + 
\frac{d\vtilde}{d\mu} \Prj \,. 
\label{dvhub}
\ee
The first term contains the bare derivative of the kernel $\Prj$ [Eq.~\eqref{dproj}]; the second, the total derivative
of the occupation matrices:
\be
\frac{d\vtilde}{d\mu}\Prj = - \sum_{Im_1m_2} U^I \, \frac{d n^{I\sigma}_{m_1m_2}}{d\mu} \, \Projj \,.
\label{dvhub2}
\ee
Based on the definition of these matrices in Eq.~\eqref{ns}, their total derivative will be also the sum of a bare and a response term:
\begin{eqnarray}
\frac{d n^{I\sigma}_{m_1m_2}}{d\mu} & = & 
\frac{\de n^{I\sigma}_{m_1m_2} }{\de \mu} \nn \\
&+& 
\sumi \biggl[ \Bra{\frac{d\psis}{d\mu}} \Proj \ket{\psis} \nn \\
& & \hspace{1cm} + \, \bra{\psis} \Proj \Ket{\frac{d\psis}{d\mu}} \biggr] \,.
\label{dns}
\end{eqnarray}
Similarly to the density response, {\it both} conduction and valence components of $\frac{d\psis(\mathbf{r})}{d\mu}$ contribute to the derivative of the occupation matrices, Eq.~\eqref{dns}.
To see this, it is convenient to rewrite the SCF response term 
in Eq.~\eqref{dns} multiplying $\frac{d\psis}{d\mu}$ by the identity $\hat{\mathcal{P}}_{c,\sigma} +\hat{\mathcal{P}}_{v,\sigma} = 1$:
\begin{eqnarray}
\frac{d n^{I\sigma}_{m_1m_2}}{d\mu} & = & \frac{\de n^{I\sigma}_{m_1m_2} }{\de \mu} \nn \\
& & + \, \sumi \Bra{\frac{d\psis}{d\mu}} \left( \hat{\mathcal{P}}^\dagger_{c,\sigma} + \hat{\mathcal{P}}^\dagger_{v,\sigma} \right) \Proj \Ket{\psis} \nn \\
& & + \sumi 
\Bra{\psis} \Proj \left( \hat{\mathcal{P}}_{c,\sigma} + \hat{\mathcal{P}}_{v,\sigma} \right) \Ket{\frac{d\psis}{d\mu}} \nn \\
& \equiv & \frac{\de n^{I\sigma}_{m_1m_2} }{\de \mu} + \Dn + \delta^{\mu} n^{I\sigma}_{m_1 m_2} \,.
\label{dns_scf1}
\end{eqnarray}
The projections on the conduction  manifold in Eq.~\eqref{dns_scf1} are conveniently
rewritten using the definition in Eq.~\eqref{projc}:
\begin{eqnarray}
\Dn & \equiv & \sumi
\left[ \bra{\Dpsis} \Proj \ket{\psis} \right. \nn \\
& & \hspace{0.8cm} + \, \left. \bra{\psis} \Proj \ket{\Dpsis} \, \right] \,,
\label{dns_cond}
\end{eqnarray}
and contain terms directly accessible from $\ket{\Dpsis}$, the solutions of Eq.~\eqref{dks2}. The projections on the valence state manifold, instead, contain only {\it bare} derivatives. To understand this, it is useful to start from the derivative of Eq.~\eqref{ortho}:
\be
\Bra{\frac{d\psis}{d\mu}} \hat{S} \Ket{\psjs} + \Bra{\psis} \hat{S} \Ket{\frac{d\psjs}{d\mu}} 
= - \Bra{\psis} \frac{\partial \hat{S}}{\partial\mu} \Ket{\psjs} \,.
\label{dpsi2}
\ee
Using this equation and the definition of  $\hat{\mathcal{P}}_{v,\sigma}$ [Eq.~\eqref{projv}] in Eq.~\eqref{dns_scf1}, the valence component $\delta^{\mu} n^{I\sigma}_{m_1 m_2}$ of the response occupation matrix (last term in Eq.~\eqref{dns_scf1}) remains determined as:
\be
 \delta^{\mu} n^{I\sigma}_{m_1 m_2} = - \sumi \, \bra{\psis} \Proj \ket{\delta^{\mu} \psis} \,,
 \label{dnsv}
\ee
where the valence component of the response KS wave functions $\ket{\delta^{\mu} \psis}$ is a short-hand notation for the following quantity:
\be
\ket{ \delta^{\mu} \psis } = \sumj^{\mathrm{occ}} \Ket{\psjs} \Bra{\psjs} \frac{\de \hat{S}}{\de\mu} \Ket{\psis} \,.
\label{dpsiorth}
\ee  
Again, it is worth noting that the bare derivative $\delta^{\mu} n^{I\sigma}_{m_1 m_2}$ stems from the use of US PPs and has no corresponding counterpart in the NC PPs scheme.

\subsection{Hubbard contributions to the matrix of force constants}
\label{sec:Hub_force_constants}

The matrix of interatomic force constants is defined as~\cite{baroni01}:
\be
C_{\mu\lambda} = \frac{d^2 E_{\mathrm{DFT}+U}}{d\mu d\lambda}
= -\frac{dF_\lambda}{d\mu} \,.
\label{eq:force_const_def}
\ee 
The expressions of the total energy $E_{\mathrm{DFT}+U}$ and of the force $F_\lambda$ are given in Eqs.~\eqref{edftu} and ~\eqref{for2}, respectively. Taking the total SCF derivative of $F_\lambda$ we obtain:
\begin{eqnarray}
C_{\mu\lambda} & = & 
\hspace{-0.1cm} \sumis 
\Bra{\psis} \frac{d}{d\mu} \left[ \frac{\de \hat{V}_\mathrm{KS}^{\sigma}}{\de \lambda} + 
\vtilde \frac{\de \Prj}{\de \lambda} - \eis \frac{\de \hat{S}}{\de\lambda} \right] \Ket{\psis} \label{dyndv}  \nn \\
&+& \sumis 
\left\{ \Bra{\frac{d\psis}{d\mu}} \left[ \frac{\de \hat{V}_\mathrm{KS}^{\sigma}} {\de \lambda} + 
\vtilde \frac{\de \Prj}{\de \lambda} - \eis \frac{\de \hat{S}}{\de\lambda} \right] \Ket{\psis} \right. \label{dyndpsi} \nn  \\
& & \hspace{1.5cm} \biggl. + \, \mathrm{c.c.}  \biggr\} \,,
\label{dyndt}
\end{eqnarray}
where c.c. indicates the complex conjugate. 

In the following we will derive only contributions stemming from the Hubbard correction; other terms are discussed in detail in Refs.~\cite{dalcorso97, dalcorso01}. The first Hubbard term comes from the first line of Eq.~\eqref{dyndv}  (second term inside the square brackets) and can be expressed as:
\begin{eqnarray}
C^{(a)}_{U,\mu\lambda} & = & \sumis 
\Bra{\psis} \frac{d}{d\mu} \left[ \vtilde \frac{\de\Prj}{\de\lambda} \right] \Ket{\psis} \nn  \\
& = & \sumis \left[ \Bra{\psis} \vtilde \frac{\de^2 \Prj}{\de\mu\de\lambda} \Ket{\psis} \right. \nn \\
& & \hspace{1.2cm} + \, \left. \Bra{\psis} \frac{d\vtilde}{d\mu} \frac{\de\Prj}{\de\lambda} \Ket{\psis} \right] \,,
\label{force_const_a}
\end{eqnarray}
where we note the presence of a second order bare derivative of the projector $\Prj$ and a ``mixed" term containing the product of a first order total derivative of $V^\sigma_U$ and a first order bare derivative of $\Prj$. 

A second Hubbard term is obtained from the second and third lines of Eq.~\eqref{dyndpsi} by projecting $\frac{d\psis}{d\mu}$ on the conduction manifold. Resolving the unity as the sum of projectors on valence and conduction states [as in Eq.~\eqref{dns_scf1}], this term, present in both US and NC PPs implementations, can be written as follows: 
\bea
\label{force_const_b}
C^{(b)}_{U,\mu\lambda} = 2 \, \mathrm{Re} \left\{\sumis 
\Bra{\Dpsis} \vtilde \frac{\de \Prj}{\de \lambda} \Ket{\psis} \right\} \,.
\eea
The remaining Hubbard contributions to the matrix of force constants stem from the projection of the response wave function (second and third lines of Eq.~\eqref{dyndpsi}) on the valence manifold and from the derivative of the energy eigenstate resulting from the first line of Eq.~\eqref{dyndpsi}. Since they are all related to the generalized orthonormality of KS wave functions, Eq.~\eqref{ortho}, (and thus specific to US PPs) these terms are conveniently treated and lumped together to achieve a more compact final expression.
Let us start from the Hubbard contribution that stems from the response of the KS eigenvalues (first line of Eq.~\eqref{dyndt}): 
\be
-\sumis \frac{d\eis}{d\mu} \Bra{\psis} \frac{\de \hat{S}}{\de\lambda} \Ket{\psis} \,.
\label{deis1}
\ee
The factor $\frac{d\eis}{d\mu}$ can be derived from Eq.~\eqref{dks} 
by projecting both sides on $\bra{\psis}$ (cf. with Eq.~(22) in Ref.~\cite{dalcorso01}):
\be
\frac{d\eis}{d\mu} = \Bra{\psis} \left[ \frac{d \hat{V}^\sigma_\mathrm{KS}}{d\mu} + \frac{d \hat{V}^\sigma_\mathrm{Hub}}{d\mu} - \eis \frac{\de \hat{S}}{\de\mu} \right] \Ket{\psis} \,.
\label{deis2}
\ee
Using Eq.~\eqref{dvhub}, the Hubbard contributions to Eq.~\eqref{deis1} (from the middle term within the parenthesis in Eq.~\eqref{deis2}) can be written as: 
\be
-\sumis
\Bra{\psis} \left[ \frac{d\vtilde}{d\mu} \Prj + \vtilde \frac{\de \Prj}{\de \mu} \right] \Ket{\psis} 
\Bra{\psis} \frac{\de \hat{S}}{\de\lambda} \Ket{\psis} \,.
\label{deis_hub}
\ee
This contribution is to be summed to Hubbard terms in the second and third lines of Eq.~\eqref{dyndpsi} coming from the projection of the response wave function on the valence manifold. These latter terms can be obtained by substituting $\Ket{\frac{d\psis}{d\mu}}$ in Eq.~\eqref{dyndpsi} with the explicit expression of $\Ket{\delta^{\mu}\psis}$,  Eq.~\eqref{dpsiorth}, and by using the following equation (valid for $j \ne i$)
\be
\Bra{\psjs} \hat{S} \Ket{\frac{d\psis}{d\mu}} = 
\frac{\Bra{\psjs} \left[ \frac{d\hat{V}^\sigma_\mathrm{KS}}{d\mu} + 
\frac{d\hat{V}^\sigma_\mathrm{Hub}}{d\mu} - \eis \frac{\partial \hat{S}}{\partial\mu} \right] 
\Ket{\psis}}{\eis - \ejs} \,,
\ee
(cf. with Eq.~(23) in Ref.~\cite{dalcorso01}), which can be  obtained by projecting both members of Eq.~\eqref{dks} on $\bra{\psjs}$. 
Summing all the response terms  of Eq.~\eqref{dyndpsi} containing the derivative of the Hubbard potential with those arising from Eq.~\eqref{deis_hub} one obtains the third and fourth Hubbard terms of the matrix of force constants: 
\begin{eqnarray}
C^{(c)}_{U,\mu\lambda} & = & - \sumis \left[ \Bra{\psis} \vtilde \frac{\de \Prj }{\de \mu} \Ket{\delta^{\lambda} \psis} \right. \nn \\
& & \hspace{1.5cm} + \, \left. \Bra{\delta^{\mu}\psis} \vtilde \frac{\de \Prj }{\de \lambda} \Ket{\psis} \right] \,, 
\label{force_const_c}
\end{eqnarray}
\bea
C^{(d)}_{U,\mu\lambda} = - \sumis \Bra{\psis} \frac{d \vtilde }{d \mu } \Prj \Ket{\delta^{\lambda} \psis} \,.  
\label{force_const_d}
\eea
These terms represent, respectively, a bare and a SCF contribution to the matrix of force constants \cite{Note_totder}.

Using the definitions introduced in Eqs.~\eqref{vtilde}, \eqref{dnbare}, \eqref{VP_shorthand}, \eqref{dvhub2}, \eqref{dns_scf1}, and \eqref{dnsv} it is now convenient to regroup the Hubbard terms of the matrix of force constants (contained in the expression of $C_{U,\mu\lambda}^{(a)}-C_{U,\mu\lambda}^{(d)}$) in three main contributions which contain, respectively, bare derivatives, orthonormality terms (specific to US PPs), and linear-response terms depending on the readjustment of the wave functions. 
The first piece comes entirely from Eq.~\eqref{force_const_a} and contains the second bare derivative and the product of first bare derivatives of the occupation matrices:
\begin{eqnarray}
C^{(1)}_{U,\mu\lambda} & = & 
\sumIsmm \vtildeImm \, \frac{\de^2 n^{I\sigma}_{m_2 m_1}}{\de\mu \de\lambda} \nn \\
& & - \sumIsmm U^I \, \frac{\de n^{I\sigma}_{m_1 m_2}}{\de\mu} 
\frac{\de n^{I\sigma}_{m_2 m_1}} {\de\lambda} \,.
\label{force_const_1}
\end{eqnarray}

The second term 
results from the summation of Eq.~\eqref{force_const_c} and the bare terms of Eqs.~\eqref{force_const_a} and \eqref{force_const_d}:
\begin{eqnarray}
C^{(2)}_{U,\mu\lambda} & = &
-\sumIsmm \vtildeImm \, \sumi \left[ \Bra{\psis} \frac{\de \Projj}{\de\mu} \Ket{\delta^\lambda \psis}  \right. \nn \\
& & \hspace{2.5cm} + \left. \Bra{\delta^\mu \psis} \frac{\de \Projj}{\de\lambda} \Ket{\psis} \right] \nn \\
& & - \sumIsmm U^I \left[ \delta^\mu n^{I\sigma}_{m_1 m_2} \frac{\de n^{I\sigma}_{m_2 m_1}}{\de\lambda} \right. \nn \\
& & \hspace{2.5cm} + \left. \delta^\lambda n^{I\sigma}_{m_1 m_2} \frac{\de n^{I\sigma}_{m_2 m_1}}{\de\mu} \right] \nn \\
& & - \sumIsmm U^I \, \delta^\mu n^{I\sigma}_{m_1 m_2} \, \delta^\lambda n^{I\sigma}_{m_2 m_1} \,.
\label{force_const_2}
\end{eqnarray}
This term is also bare and contains all the pieces resulting from the generalized orthonormality conditions, Eq.~\eqref{ortho}.
Finally, from Eqs.~\eqref{force_const_a}, \eqref{force_const_b}, and \eqref{force_const_d} one can collect in a third SCF term all contributions that depend on the KS states response:
\begin{eqnarray}
\label{force_const_3}
C^{(3)}_{U,\mu\lambda} & = &
2 \, \mathrm{Re} \left\{ \sumIsmm \vtildeImm \right. \times \nn \\
& & \hspace{1.8cm} \times \left. \sumi \Bra{\Dpsis} \frac{\de \Projj}{\de\lambda} \Ket{\psis} \right\} \nn \\
& & - \sumIsmm U^I \, \Dn \left( \frac{\de n^{I\sigma}_{m_2 m_1}}{\de\lambda} + \delta^\lambda n^{I\sigma}_{m_2 m_1} \right) \,. \nn \\
& &
\end{eqnarray}
It is important to note that 
the derivatives of any order of the Hubbard $U$ are neglected here (as in most works in literature). This is obviously an approximation whose validity should be tested carefully, case by case \cite{Note_Ud}.

\subsection{Hubbard contribution to the nonanalytic part of the dynamical matrix in polar insulators}
\label{polar}

Another important contribution from the Hubbard correction to the dynamical matrix (i.e. the Fourier transform of the matrix of force constants, normalized by the square-root of atomic masses), is in the dynamical matrix nonanalytic part $\tilde{C}_{\mu\lambda}(\mathbf{q})$~\cite{baroni01, giannozzi91}. This results from the coupling of longitudinal vibrations with the macroscopic electric field generated by the displacements of ions, and is responsible for the splitting in energy between 
the longitudinal and transverse optical modes (LO-TO splitting) of ionic semiconductors and insulators~\cite{cochran}.
The correction to fully account for this coupling needs to be
computed and added separately only at $\vctr{q}=\mathbf{0}$ (the response at finite $\vctr{q}$ vectors automatically accounts for these effects).  
It can be shown that 
$\tilde{C}_{\mu\lambda}(\mathbf{q})$ is a function of the 
Born effective charges tensor $\vctr{Z}^*$ and the high-frequency 
(electronic) dielectric tensor $\boldsymbol{\epsilon}^{\infty}$~\cite{cochran}:
\be
\tilde{C}^{\alpha\beta}_{IJ}(\mathbf{q}) = 
\frac{4\pi}{\Omega}
\frac{(\vctr{q}\cdot\vctr{Z}^*_I)_{\alpha} (\vctr{q}\cdot\vctr{Z}^*_J)_{\beta}}
{\vctr{q}\cdot \boldsymbol{\epsilon}^{\infty} \cdot\vctr{q}} \,,
\label{dyn_matrix_na}
\ee
where $I$ and $J$ are  atomic indices, and $\alpha$ and $\beta$ are Cartesian components. 
The calculation of the $\boldsymbol{\epsilon}^{\infty}$ tensor is based on the response of the electronic system to a macroscopic electric field~\cite{baroni86,giannozzi91}, and requires the expectation value of the corresponding electrostatic potential (proportional to the position operator $\hat{\vr}$) between conduction and valence KS states.  
This is evaluated via the expression~\cite{tobik04}:
\bea
\bra{ \psi_{c\sigma} } \hat{S} \hat{\vr} \ket{\psi_{v\sigma} }  = 
\frac{ \bra{\psi_{c\sigma} } \, [\hat{H}^\sigma - \varepsilon_{v\sigma}\hat{S}, \hat{\vr}] \, \ket{\psi_{v\sigma}}} 
{\varepsilon_{c\sigma} - \varepsilon_{v\sigma} } \,,
\label{eq:Hr_comm}
\eea
where $c$ and $v$ refer to the conduction- and valence-states manifolds, respectively, and $\hat{H}^\sigma$ is the single-particle Hamiltonian defined in Eq.~\eqref{eq:Hamiltonian}. Besides the kinetic operator in $\hat{H}^\sigma$, only non-local components of the potential $\hat{V}^{\sigma}$ [Eq.~\eqref{pot}] contribute to the commutator $[\hat{H}^\sigma, \hat{\vr}]$.
Due to its non-local nature, the Hubbard potential $\hat{V}_\mathrm{Hub}^{\sigma}$ [Eq.~\eqref{vhub}] contributes to this quantity via the following term:
\begin{eqnarray}
& & \bra{\psi_{c\sigma}} \, [\hat{V}_\mathrm{Hub}^{\sigma}, \hat{\vr}] \, \ket{\psi_{v\sigma}} = \nonumber \\
& & \hspace{0.35cm} \sum_{I m_1 m_2} V_{U, m_1 m_2}^{I \sigma} \bra{\psi_{c\sigma}} \, \left[\hat{S} \ket{\f^I_{m_1}}\bra{\f^I_{m_2}} \hat{S}, \hat{\vr}\right] \, \ket{\psi_{v\sigma}} \,,
\label{vhubcomm}
\end{eqnarray}
where $V_{U, m_1 m_2}^{I \sigma}$ is defined in Eq.~\eqref{vtilde}. This quantity can be conveniently evaluated in reciprocal space, as shown in Ref. ~\cite{dalcorso01} for analogous terms.

The Born effective charges tensor $\vctr{Z}^*$ can be evaluated as a mixed second derivative of the total energy~$E_{\mathrm{DFT}+U}$~\cite{Ghosez:1995}: 
\be
Z^{*}_{J\alpha\beta} = - \frac{d}{dE_{\alpha}} \left( \frac{d E_{\mathrm{DFT}+U}}{d u_{J\beta}} \right) =  \frac{d F_{J\beta}}{dE_{\alpha}} \,,
\label{zss}
\ee
where $E_{\alpha}$ is the $\alpha$ component of the macroscopic electric field, $u_{J\beta}$ is the $\beta$ component of the $J$-th atom displacement, and $F_{J\beta}$ is the corresponding force [Eq.~\eqref{for2}].  
In the following, we set $\lambda = u_{J\beta}$ for notational consistency with previous sections. Considering Eqs.~\eqref{for2} and \eqref{zss}, $\vctr{Z}^*$ takes the form: 
\begin{eqnarray}
 Z^*_{\alpha\lambda} & = & -\frac{d}{dE_{\alpha}} \left( \sumis 
 \Bra{\psis} \left[ \frac{\de \hat{V}_\mathrm{KS}^{\sigma}}{\de \lambda} \right. \right. \nonumber \\
 & & \hspace{1.3cm} \left. \left. + \vtilde \frac{\de \Prj }{\de\lambda} - \eis \frac{\de \hat{S}}{\de\lambda} \right] \Ket{\psis}\right) \,,
 \label{zs}
\end{eqnarray}
where $\vtilde \frac{\de \Prj }{\de \lambda}$ is defined in Eq.~\eqref{VP_shorthand}. We will have then: 
\bea
Z^*_{\alpha\lambda} & = & - \sumis 
 \Bra{\psis} \frac{d}{dE_{\alpha}} \left[ \frac{\de \hat{V}_\mathrm{KS}^{\sigma}}{\de \lambda} + \vtilde \frac{\de \Prj }{\de\lambda} - \eis \frac{\de \hat{S}}{\de\lambda} \right] \Ket{\psis} \nonumber \\
&-& \sumis \Biggl\{
\Bra{\frac{d\psis}{dE_{\alpha}}}  \left[ \frac{\de \hat{V}_\mathrm{KS}^{\sigma}}{\de \lambda} + \vtilde \frac{\de \Prj }{\de\lambda} - \eis \frac{\de \hat{S}}{\de\lambda} \right] \Ket{\psis} \Biggr. \nonumber \\ 
& & \hspace{1cm} \Biggl. + \, \mathrm{c.c.} \Biggr\} \,.
\label{zsl_tot0}
\eea
From now on, we will consider only Hubbard contributions to $\vctr{Z}^*$  in Eq.~\eqref{zsl_tot0}. The first term comes from the first line 
(second term inside the square brackets) which, using Eqs.~\eqref{vtilde}, \eqref{dnbare}, and \eqref{VP_shorthand}, can be expressed as:
\begin{eqnarray}
Z^{* (a)}_{U,\alpha\lambda} & = & - \sumis 
 \Bra{\psis} \frac{d}{dE_{\alpha}} \left[ \vtilde \frac{\de \Prj }{\de\lambda} \right] \Ket{\psis} \nonumber \\ [5pt]
 & = & \sum_{I \sigma m_1 m_2}  U^I  \frac{d n^{I\sigma}_{m_1m_2}}{dE_{\alpha}} \frac{\de n^{I\sigma}_{m_2 m_1}}{\de\lambda} \,,
 \label{zsl_a}
\end{eqnarray}
where the bare term $\frac{\de n^{I\sigma}_{m_2 m_1} }{\de\lambda}$ is defined in Eq.~\eqref{dnbare}.
The second Hubbard term is obtained from the last term 
of the first line in Eq.~\eqref{zsl_tot0}:
\bea
\sumis \frac{d\eis}{dE_{\alpha}} \Bra{\psis}   \frac{\de \hat{S}}{\de\lambda}\Ket{\psis} \,,
\label{zsl}
\eea
which is similar to the one in Eq.~\eqref{deis1}, the difference being that here the KS eigenvalues respond to the macroscopic electric field rather than to atomic displacements. Similarly to Eq.~\eqref{deis_hub}, this term reads
\begin{eqnarray}
 Z^{* (b)}_{U,\alpha\lambda} &= &  \sumis \Bra{\psis}  \frac{d\vtilde}{dE_{\alpha}} \Prj  \Ket{\psis} 
 \Bra{\psis} \frac{\de \hat{S}}{\de\lambda} \Ket{\psis} \nonumber \\
 & = & -\sum_{I m_1 m_2 i\sigma}  U^I  \frac{d n^{I\sigma}_{m_1m_2}}{dE_{\alpha}} \Bra{\psis} \Projj \Ket{\psis} \nonumber \\
 & & \hspace{1.5cm} \times \, \Bra{\psis} \frac{\de \hat{S}}{\de\lambda}\Ket{\psis} \,,
 \label{zsl_b} 
\end{eqnarray}
where we have used Eq.~\eqref{vtilde}. Note that in the derivation of Eqs.~\eqref{zsl_a} and \eqref{zsl_b} we used the fact that only the KS wave functions respond to the macroscopic electric field (nuclei are clamped),  therefore bare terms as $\frac{d}{dE_{\alpha}} \frac{\de \Prj }{\de\lambda}$ and $\frac{d}{dE_{\alpha}} \frac{\de \hat{S}}{\de\lambda}$ are zero. The derivative $\frac{d n^{I\sigma}_{m_1m_2}}{dE_{\alpha}}$, appearing in Eqs.~\eqref{zsl_a} and \eqref{zsl_b}, is a pure SCF response term and it is computed as:
\begin{eqnarray}
\frac{d n^{I\sigma}_{m_1m_2}}{dE_{\alpha}} & = & \sumi \biggl[ \Bra{\frac{d\psis}{dE_{\alpha}}} \Proj \ket{\psis} \biggr. \nonumber \\
& & \hspace{1cm} \biggl. + \, \bra{\psis} \Proj \Ket{\frac{d\psis}{ dE_{\alpha}   }} \biggr] \,.
\label{dnse} 
\end{eqnarray}
Similarly to Eq.~\eqref{dns_scf1}, we can insert an identity in the equation above and obtain the conduction and valence components of the response occupation matrix~$\frac{d n^{I\sigma}_{m_1m_2}}{dE_{\alpha}}$. The conduction component is computed as:
\begin{eqnarray}
\DnE & \equiv & \sumi
\left[ \bra{\DpsisE} \Proj \ket{\psis} \right. \nn \\
& & \hspace{0.8cm} + \, \left. \bra{\psis} \Proj \ket{\DpsisE} \, \right] \,,
\label{dns_cond_E}
\end{eqnarray}
where $\DpsisE$ is the conduction component of the response KS wave functions. It is computed by solving a Sternheimer equation analog to Eq.~\eqref{dks2} ~\cite{tobik04,baroni01} but including an electrostatic perturbing potential, calculated using Eqs.~\eqref{eq:Hr_comm} and~\eqref{vhubcomm}. Note that, as electric field perturbations involve pure SCF derivatives, bare terms analog to  the ones in Eqs.~\eqref{dnbare}, \eqref{dnsv}  and \eqref{dpsiorth} are zero, and hence the valence component of $\frac{d n^{I\sigma}_{m_1m_2}}{dE_{\alpha}}$ is zero.

The third Hubbard term comes from the second and third lines in Eq.~\eqref{zsl_tot0}:
\begin{eqnarray}
 Z^{* (c)}_{U,\alpha\lambda} & = & -\sumis \left\{
\Bra{\frac{d\psis}{dE_{\alpha}}} \vtilde \frac{\de \Prj}{\de\lambda} \Ket{\psis} + \mathrm{c.c.} \right\} \nonumber \\
& = & - \sum_{I m_1 m_2 i\sigma} U^I \left(\frac{\delta_{m_1 m_2}}{2} - n^{I\sigma}_{m_1 m_2}\right) \nonumber \\
& & \hspace{0.5cm} \times \left\{ \Bra{\frac{d\psis}{dE_{\alpha}}} \frac{\de \Projj}{\de\lambda} \Ket{\psis} + \mathrm{c.c.} \right\} \,,
\label{zsl_c}
\end{eqnarray}
where we have used Eq.~\eqref{vtilde}. Therefore, the three Hubbard contributions to the $Z^*$ Born effective charges are given by Eqs.~\eqref{zsl_a}, \eqref{zsl_b}, and \eqref{zsl_c}. It is possible to rewrite $Z^{* (a)}_{U,\alpha\lambda}$ and $Z^{* (c)}_{U,\alpha\lambda}$ as one compact term, by performing manipulations on Eq.~\eqref{zsl_a} using Eqs.~\eqref{dnbare} and \eqref{dnse}. By doing so, and by including $Z^{* (b)}_{U,\alpha\lambda}$, we can write the final expression for all Hubbard contributions to $Z^*$ as:
\begin{eqnarray}
Z^{*}_{U,\alpha\lambda} & = & -\sumis \left\{
\Bra{\frac{d\psis}{dE_{\alpha}}} \frac{\de \hat{V}^{\sigma}_\mathrm{Hub}}{\de\lambda} \Ket{\psis} + \mathrm{c.c.} \right\} \nonumber \\
& & + \sumis \Bra{\psis}  \frac{d\vtilde}{dE_{\alpha}} \Prj  \Ket{\psis} 
 \Bra{\psis} \frac{\de \hat{S}}{\de\lambda} \Ket{\psis} \,,
\label{zsl_tot}
\end{eqnarray}
where
\bea
\frac{\de \hat{V}^{\sigma}_\mathrm{Hub}}{\de\lambda} = \vtilde \frac{\de \Prj}{\de\lambda} + 
\frac{\de \vtilde}{\de\lambda} \Prj 
\label{dvhubb}
\eea
is the bare derivative of the Hubbard potential [Eq.~\eqref{vhub}].

An alternative (but equivalent) way to compute $Z^{*}_{U,\alpha\lambda}$ can be defined by
changing the order of derivatives in Eq.~\eqref{zss}. This alternative procedure was also  implemented and has been proved to give identical results to the one derived above from Eq.~\eqref{zss}. Further benchmarks of the nonanalytic terms of the dynamical matrix are contained in the SM~\cite{SupplementalMaterial}, 
Sec.~\ref{secSM:nonanalytic_term}.

\section{Technical details}
\label{sec:technical_details}

The DFPT+$U$ approach has been implemented in the
\textsc{Quantum ESPRESSO} distribution~\cite{Giannozzi:2009, Giannozzi:2017}
and it is publicly available. 
Calculations are performed using the plane-wave (PW) pseudopotential method and the generalized-gradient approximation (GGA) for the xc functional constructed with the PBEsol prescription~\cite{Perdew:2008}. US PPs were taken from the GBRV library~\cite{Garrity:2014, Note_USPP}. Kohn-Sham wave functions and charge density were expanded in PWs up to a kinetic-energy cutoff of 60~Ry and 720~Ry, respectively. 

Electronic ground states were computed by sampling the Brillouin zones (BZ) with uniform, $\Gamma$-centered $6 \times 6 \times 6$ and $4 \times 4 \times 4$ $\mathbf{k}$-point meshes for CoO and LiCoO$_2$, respectively. The CoO GGA metallic ground state required instead a $18 \times 18 \times 18$ $\mathbf{k}$-point grid. For phonon calculations the  LiCoO$_2$ $\mathbf{k}$-point grid  was refined to $8 \times 8 \times 8$. Given the non-magnetic nature of its ground state, the KS states occupations were held fixed for LiCoO$_2$. 
For CoO, instead, a gaussian smearing  was used, except for GGA calculations that, owing to the metallic character of the ground state, was better treated by a Marzari-Vanderbilt smearing~\cite{marzari99}. In all cases a broadening width of 0.01 Ry was adopted. 

The effective Hubbard $U$ was determined from first-principles using the DFPT approach of Ref.~\cite{timrov18}, with  uniform $4 \times 4 \times 4$ and $3 \times 3 \times 3$ $\mathbf{q}$-point meshes for CoO and LiCoO$_2$, respectively.
The Hubbard $U$ was computed (typically with a precision of about 0.01~eV) in a self-consistent way, i.e. cycling linear-response calculations of $U$~\cite{timrov18} and structural optimizations  
until convergence, 
as explained in Refs.~\cite{Hsu:2009,himmetoglu14}.  
For the Co $3d$ states of CoO and LiCoO$_2$ we obtained, respectively, $U_\mathrm{scf}=$~4.55~eV and $U_\mathrm{scf}=$~6.91~eV.

DFPT and DFPT+$U$ phonon calculations  were performed using  optimized crystal structures, with a uniform, $\Gamma$-centered $4 \times 4 \times 4$
$\mathbf{q}$-point mesh.

The projectors of Eq.~\eqref{proj} are constructed using  atomic orbitals present in the pseudopotentials. 
The linear-response KS equations~\eqref{dks2} were solved using the conjugate-gradient algorithm~\cite{Payne:1992} and the mixing scheme of Ref.~\cite{Johnson:1988} for the Hxc potential response to speed up convergence. The labeling of the BZ high-symmetry points and directions for the phonon dispersion were determined using SeeK-path~\cite{Hinuma:2017}. 
Phonon analysis was made with the help of the Materials Cloud Interactive Phonon visualizer~\cite{PHvis}.
The data used to produce the results of this work are available on the Materials Cloud Archive~\cite{MaterialsCloudData}.

\section{Results}

\label{validation}

We now showcase the importance of the Hubbard correction for the calculation of the vibrational spectrum of two paradigmatic transition-metal oxides, CoO and LiCoO$_2$. We briefly discuss also  their equilibrium crystal structures and electronic band gaps. 

\subsection{CoO}

Prototypical representative of correlated transition-metal monoxides (TMO), CoO is 
investigated for a number of technological applications including spintronics~\cite{Raquet:1998}, gas-sensing~\cite{Wollenstein:2003}, photo-catalysis  
~\cite{demkov14}
and energy storage~\cite{Poizot:2000, Ruffo:2016}. 
 As other TMOs, in its high-temperature paramagnetic phase CoO has a rock-salt type structure (space group $Fm3m$) with a lattice parameter of 4.26~\AA~\cite{Jauch:2001}.
A phase transition to a type II antiferromagnetic (AFII) ground state takes place below the N\'eel  temperature  $T_\mathrm{N}\approx 291$~K. In this phase the material presents ferromagnetic (111) planes stacked antiferromagnetically along the [111] direction~\cite{shull51,roth58}. This magnetic order is compatible with rhombohedral symmetry.
However, at variance with other TMOs, CoO does not exhibit a simple rhombohedral deformation, but it rather adopts  a monoclinic symmetry (space group $C2/m$)~\cite{Jauch:2001}. 
The latter can be seen as resulting from a rhombohedral distortion of a tetragonal unit cell characterized by a slightly contracted cubic $c$-axis ($c/a=0.988$)~\cite{Jauch:2001,Dalverny:2010,Wdowik:2008,kant08}.

Notwithstanding this rather complex structural behavior, in order to simplify the material's description (avoiding the effects of the rather small monoclinic deformation) and  contain computational costs, 
our phonon calculations were performed on a four-atoms AFII rhombohedral unit cell
rather than on the larger, eight-atoms monoclinic one. This choice is justified by a comparative analysis 
of the energies and equilibrium crystal structures obtained from the monoclinic and the rhombohedral  cells.  Table~\ref{table:CoO_lat_param} shows that the two structures are practically 
indistinguishable (0.05\% difference in the lattice parameters) and that their energies are within 2.0 and 0.4 meV per formula unit in GGA and GGA+$U$, respectively, which are comparable with the precision of our calculations. A clear evidence emerging from Table~\ref{table:CoO_lat_param} is, instead, the net improvement the Hubbard correction brings to
the crystal structure
(with respect to GGA), in excellent  agreement with low-temperature experimental  data~\cite{Jauch:2001}.
This includes also a marginal stabilization of the monoclinic structure (less stable in GGA).
\begin{table}[t]
\begin{center}
  \begin{tabular}{lcccccc}
    \hline\hline
&   \parbox{0.8cm}{$a$ (\AA)}  & \parbox{0.8cm}{$b$ (\AA)} & \parbox{0.8cm}{$c$ (\AA)} &  \parbox{0.8cm}{$\beta$ ($^\circ$)} &   \parbox{1.6cm}{$\Delta E$ (meV/f.u.)}& \parbox{1.2cm}{$E_g$ (eV)}      \\ \hline
GGA$^R$ &  5.577 &  2.748 & 3.109 &  132.78   & 0  &  0                              \\
GGA$^M$ &  5.575 &  2.749 & 3.108 &  132.74   & 2.0    &  0                                 \\
GGA+$U^R$  & 5.206  & 3.019 & 3.009 & 125.05  & 0.4  &  2.32                                 \\
GGA+$U^M$  & 5.209  & 3.018 & 3.010 & 125.08  &  0    &  2.32                               \\
Expt.  & 5.182 & 3.018  & 3.019 &  125.58     &      & 2.5 $\pm$ 0.3  \\
\hline\hline
   \end{tabular}
\end{center}
\caption{CoO experimental and theoretical lattice parameters ($a$, $b$, $c$, and $\beta$) of the monoclinic crystal, and energy differences from the most stable structure ($\Delta E=0$), as obtained from calculations in the rhombohedral ($^R$) and monoclinic ($^M$) unit cells for both GGA and GGA+$U$.  The calculated and experimental band gaps ($E_g$) are also shown.  Experimental lattice parameters are from Ref.~\cite{Jauch:2001}, 
the band gap from Ref.~\cite{vanElp:1991}. 
}
\label{table:CoO_lat_param}
\end{table}
The looser  agreement  with the experiment  
that GGA's structure
shows is probably related to the metallic character of its ground state (in contradiction with the observed insulating behavior of the material) that pulls Co ions closer to each other along the monoclinic $b$ direction in order to increase bands dispersion.
In fact, the GGA functional fails to localize the 
Co $3d$ electrons as a consequence of a large self-interaction error.
By reducing 
this error and favoring the electronic localization,  
the Hubbard correction stabilizes
an insulating ground state, whose improved electronic structure refines the equilibrium geometry of the crystal.

Our calculated GGA+$U$ band gap is 2.32 eV, 
which  agrees well with photoemission experiments (2.5 $\pm$ 0.3~eV)~\cite{vanElp:1991}.
We stress that this overall very good agreement with the experiments is obtained without any fitting 
parameter. 
Our Hubbard $U$ value  (4.55~eV) was determined {\it ab initio} and self-consistently using a recently-developed DFPT-based linear-response technique ~\cite{timrov18}.

Table~\ref{table:CoO_epsilon_and_z} reports
the computed high-frequency dielectric tensor and Born effective charges, in comparison with the experimental values. These quantities, obtained  through the response to a finite electric field (see Sec.~\ref{polar}), are crucial to evaluate the nonanalytic part of the dynamical matrix for polar insulators, and  to capture the LO-TO splittings of phonon frequencies  at zone center. 
While the computed dielectric tensor overestimates the experimental one by 16\%,  a very good  agreement 
is found for the Born effective charges (9\% difference between theory and experiment), with calculations suggesting a slightly more ionic character than the experiments. Note that the deviation from the cubic symmetry (due to the rhombohedral cell adopted here) results in nonvanishing off-diagonal elements of these tensors, also reported in Table~\ref{table:CoO_epsilon_and_z}. 
\begin{table}[h]
\begin{center}
  \begin{tabular}{ccc}
    \hline\hline
                               &  \parbox{1.75cm}{DFPT+$U$} &  \parbox{1.75cm}{Expt.}   \\ \hline
    $\epsilon^\infty_{ii}$     &                     6.17                &  \multirow{2}{*}{5.3}    \\ [1pt]
    $\epsilon^\infty_{ij}$     &                    -0.04                &                         \\ \hline
    $Z^*_{\mathrm{Co},ii}$     &                     2.25                &  \multirow{2}{*}{2.06}  \\ [1pt]
    $Z^*_{\mathrm{Co},ij}$     &                     0.05                &                         \\ \hline
    $Z^*_{\mathrm{O},ii}$      &                    -2.25                &  \multirow{2}{*}{-2.06} \\ [1pt]
    $Z^*_{\mathrm{O},ij}$      &                    -0.05                &                         \\ 
    \hline\hline
  \end{tabular}
\end{center}
\caption{CoO high-frequency dielectric tensor and Born effective charges  
as computed using DFPT+$U$ and measured in experiments. 
Both tensors are reported in the Cartesian framework. Diagonal and off-diagonal elements are labeled as $ii$ and $ij$. The experimental values for $\vctr{Z}^*$ and $\boldsymbol{\epsilon}^{\infty}$ are from Refs.~\cite{Wdowik:2007} and \cite{Sakurai:1968}, respectively.
}
\label{table:CoO_epsilon_and_z}
\end{table}

\begin{figure}[t]
\begin{center}
\includegraphics[width=0.49\textwidth]{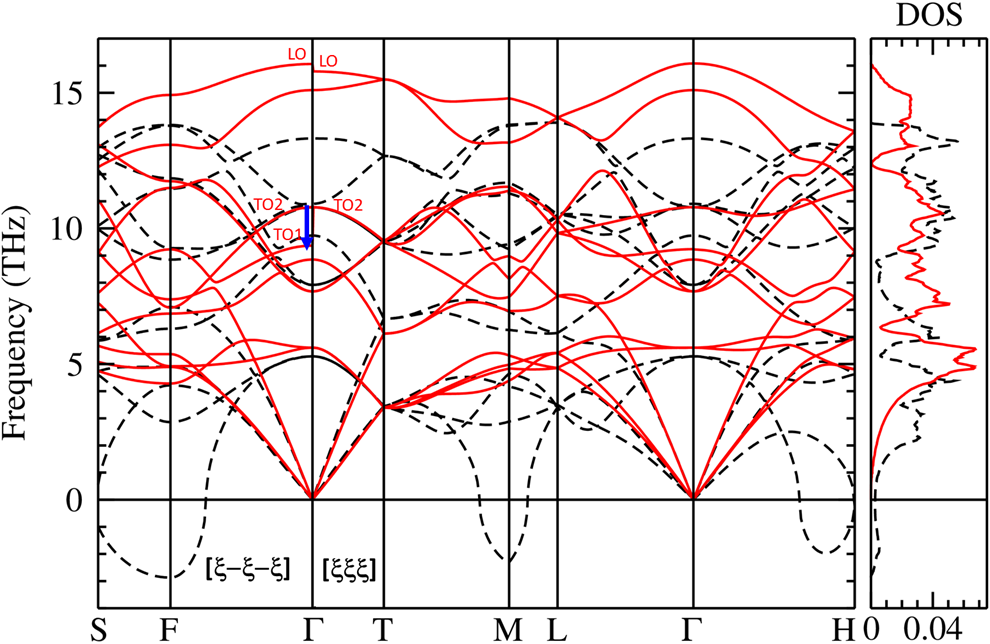}\caption{CoO phonon dispersion in THz (left panel) and phonon density of states in states/THz/cell (right panel) obtained from DFPT (black dashed lines) and DFPT+$U$ (red solid lines).  ``LO" and ``TO" label longitudinal and transverse optical modes respectively, that split at zone center. The blue arrow indicates a further  splitting between TO modes due to the AFII order (see main text). The distances along high-symmetry directions  were rescaled in the DFPT case to match the DFPT+$U$ (different) Brillouin zone.}
\label{fig:CO_disp}
\end{center}
\end{figure}

The CoO phonon dispersion, the main result of this section, is shown in Fig.~\ref{fig:CO_disp}, comparing the vibrational spectrum and density of states (DOS) obtained with DFPT (GGA functional) and DFPT+$U$ (GGA+$U$ functional). Unlike in MnO and NiO~\cite{floris11}, the difference between the two sets of results is qualitative.
The GGA functional, stabilizing a metallic ground state, predicts the rhombohedral crystal to be dynamically unstable, in agreement with what was reported in other works~\cite{Wdowik:2007, Wdowik:2008}. This is evinced from the broad negative branches (imaginary frequencies) along the S--F and F--$\Gamma$ directions, around M and along $\Gamma$--H, most of which appear to stem from optical modes. Remarkably, these instabilities are completely removed by the Hubbard correction that yields positive branches (real vibrational frequencies) across the whole BZ. Thus, the GGA+$U$ insulating crystal structure is not only in much closer agreement with experiments (Table \ref{table:CoO_lat_param}) but also dynamically stable. Looking at the phonon DOS, the GGA+$U$ frequencies are in general
blue-shifted relative to GGA (although not for all modes and not uniformly). This is true for both the low (Co-dominated) and high (O-dominated) parts of the spectrum and also leads to a more uniform distribution of the vibrational modes across the whole energy spectrum.
As a direct consequence of the insulating ground state achieved with GGA+$U$ and of the polar nature of the crystal, we have the presence  of LO-TO splittings around zone-center. These, in general, differ depending on the specific direction along which the vibration wave vector approaches $\Gamma$. Focussing, e.g., on the F--$\Gamma$--T section, the splitting is between transverse modes (denoted TO1, TO2) at 9--11~THz in Fig.~\ref{fig:CO_disp} and the longitudinal  ones (LO) 
at 16~THz.
Along F-$\Gamma$ a splitting appears between TO modes, as indicated by the blue arrow in the figure.
This splitting originates from the AFII magnetic order
that breaks the equivalence between the [111] direction ($\Gamma$-T), orthogonal to the ferromagnetic (FM) planes 
and other cubic diagonals as, e.g., the [1$\bar 1$$\bar1$] direction (F--$\Gamma$). The transverse optical modes TO1 and TO2 are degenerate along $\Gamma$-T, as they are both polarized along the FM planes. However, along F--$\Gamma$, while TO2 vibrates still on these planes (along the common orthogonal to both directions) and maintains its frequency unchanged across $\Gamma$, TO1 vibrates out of the FM planes and its frequency splits downward from that of TO2 along F--$\Gamma$.
Originally predicted by Massidda and coworkers~\cite{massidda}, these splittings between TO modes were measured experimentally (e.g., in Ref.~\cite{Chung:2003}) and also confirmed to stem from the magnetic order in previous works~\cite{floris11,Luo:2007}. 
For MnO and NiO, the   GGA+$U$  resulted in a contraction of the TO splittings with respect to GGA~\cite{floris11}, due to the reduction of the inter-site magnetic couplings following from the more pronounced localization of $3d$ electrons. The width of the splitting we obtain for CoO is much larger than for MnO and NiO, suggesting more robust magnetic couplings between Co ions than between Mn or Ni.

Note that also the topmost LO modes (at ~16 THz) exhibit a discontinuity at $\Gamma$, although
less pronounced. This small jump is also a consequence of the AFII order (the two directions $\Gamma$-T and $\Gamma$-F would otherwise be equivalent in a perfectly cubic crystal). However, it only appears because of the re-polarization of the phonons, LO along both directions, imposed by the electric field associated with the vibrations. 

We remark that the splitting of modes discussed above also contains a structural component, due to the rhombohedral distortion of the crystal and the departure from the cubic symmetry (see discussion below). 

From Fig.~\ref{fig:CO_disp} it is evident that 
the splittings obtained with GGA+$U$ are far bigger that with GGA. In fact, the GGA metallic ground state screens the macroscopic electric fields associated with LO vibrations 
and all the nonanalytic terms of the dynamical matrix vanish. The residual splittings surviving in GGA among the three TO and LO modes (at around 10.8\,THz, they are hardly visible in the figure as their amplitude is 0.13\,THz) are actually a consequence of the AFII magnetic order and of crystal distortion. Their magnetic component is 
greatly attenuated by the metallic   state while  their distortion component is
more pronounced than in GGA+$U$. Separate GGA calculations on an undistorted (rock-salt) CoO cell 
have shown that magnetic and distortion splittings 
have opposite signs, thus further reducing the overall (net) effect.

\begin{figure}[h]
\begin{center}
   \includegraphics[width=0.48\textwidth]{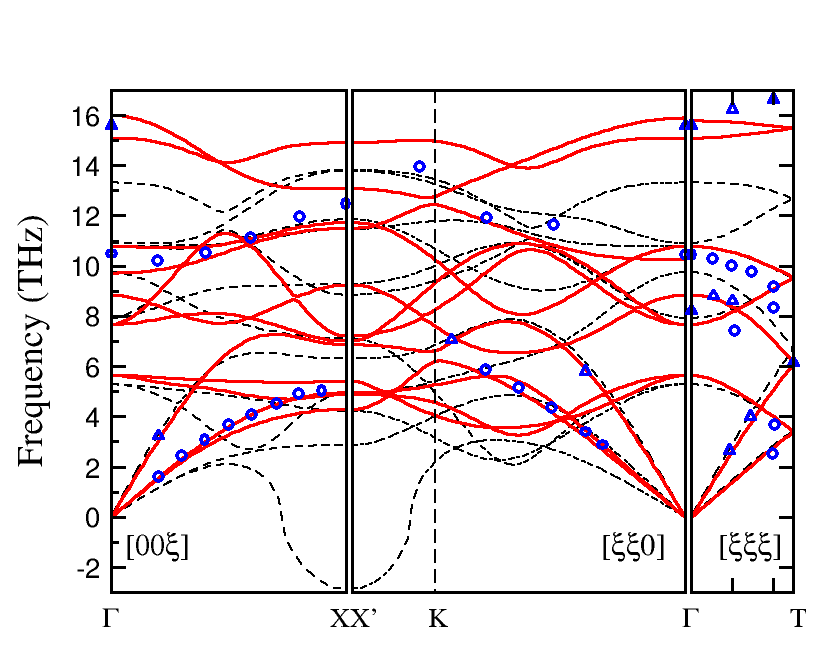}
   \caption{CoO phonon dispersion in an equivalent (distorted) two-atoms rock-salt cell compared to the experiment~\cite{Sakurai:1968}. The directions refer to  the rock-salt cell: [001] ($\Gamma$-X), [110] ($\Gamma$-K) and [111] ($\Gamma$-T). This dispersion was obtained from calculations in the four-atom rhombohedral cell of Fig.~\ref{fig:CO_disp}. Experimental data along the cubic $\Gamma$-L direction were folded to account for the doubled periodicity of the AFII cell along the [111] direction (see text). Circles (triangles) indicate transverse (longitudinal) modes. .
   }
\label{fig:LCO_disp_nacl}
\end{center}
\end{figure}

Fig.~\ref{fig:LCO_disp_nacl} shows a comparison between the phonon dispersions obtained with DFPT and DFPT+$U$ 
and data from inelastic neutron scattering (INS) experiments~\cite{Sakurai:1968}. 
Although the 
latter were performed below the Ne\'el temperature, 
the data were plotted along the high-symmetry paths of a cubic rock-salt) Brillouin zone, typical of the paramagnetic phase. Thus, theoretical phonons of the AFII cell were recomputed along these paths to facilitate the comparison, except for the [111] ($\Gamma$--T) direction along which experimental points have been folded  to account for the doubled periodicity of the AFII cell.
Clearly, doubling the number of atoms in the 
cell implies, for unfolded directions, six extra  branches compared to the experimental ones. Consistently with Ref.~\cite{Sakurai:1968}, the dispersion along the [110] ($\Sigma$) direction is 
extended beyond the border zone at K 
to reach X$^\prime$, equivalent to X.

The agreement between  
DFPT+$U$  and the experimental data is remarkably good along all directions. The GGA results, instead, miss the experimental frequencies (except for few acoustic branches), even in regions of positive phonons.
Note that the  LO-TO splittings in DFPT+$U$
improve significantly the agreement: the highest  LO and TO  modes at $\Gamma$ are matched quite well by DFPT+$U$,
while DFPT is off by non-negligible amounts (in fact more than 2 THz lower for the topmost LO mode).

Overall, our DFPT+$U$ phonon dispersion of CoO is in good agreement also with other calculations~\cite{Wdowik:2007, Wdowik:2008}. We remark, however, that the effects of the Hubbard correction on the CoO vibrational properties have been previously investigated using a direct method (based on supercells) and an empirically determined value of  $U$~\cite{Wdowik:2007,Wdowik:2008}. The results presented here are instead obtained from our DFPT+$U$ implementation 
and are based on an {\it ab initio}, linear-response evaluation of $U$ and self-consistent optimization of the crystal structure~\cite{cococcioni05,himmetoglu14,timrov18}. Further, unlike in Refs.~\cite{Wdowik:2007, Wdowik:2008} 
where  LO-TO splittings were partly obtained from experimental  $\vctr{Z}^*$ and $\boldsymbol{\epsilon}^{\infty}$, our results are 
entirely from first principles.  
Finally,  Refs.~\cite{Wdowik:2007, Wdowik:2008} used a cubic rock-salt cell, while we adopted a rhombohedral 
cell fully accounting for all geometrical deformations consistent with this symmetry.

\subsection{LiCoO$_2$}

Li$_x$CoO$_2$ is one of the most widely used cathode materials for Li-ion batteries~\cite{Miz80}.  
In its fully lithiated ($x = 1$) phase it is a non-magnetic semiconductor which crystallizes in the rhombohedral cell (containing 4 atoms) with space group $R\bar{3}m$. Table~\ref{table:LCO_lat_param} compares the equilibrium structural parameters and band gaps computed with GGA and GGA+$U$ with their experimental values. As for CoO, the GGA+$U$ results are obtained 
by a procedure where the crystal structure is reoptimized at each calculation of the Hubbard $U$,  
until the variations of both are 
within given thresholds. It is immediate to realize that 
the self-consistent $U$ correction improves significantly the agreement with experiments compared to GGA.
\begin{table}[h]
\begin{center}
  \begin{tabular}{lcccc}
    \hline\hline
                                  &   \parbox{1.4cm}{$a$ (\AA)}   &  \parbox{1.4cm}{$\alpha$ (deg)}  &  \parbox{1.4cm}{$z$} & $E_g$ (eV)
                                  \\ \hline
       GGA                        &  4.88  &  33.48  &  0.2391  & 1.14   \\
       GGA+$U$       &  4.93  &  33.10  &  0.2398  & 2.90    \\
       Expt.                      &  4.96  &  32.99  &  0.2395  & $2.7 \pm 0.3$  \\
    \hline\hline
   \end{tabular}
\end{center}
\caption{LiCoO$_2$ experimental  and theoretical lattice parameters with a rhombohedral cell and space group $R\bar{3}m$. $a$ is the lattice constant, $\alpha$ is the rhombohedral angle, $z$ is the internal (dimensionless) positional parameter of O atoms along the trigonal axis, and $E_g$ is the band gap. Experimental lattice parameters are from Ref.~\cite{Akimoto:1998},  the band gap from Ref.~\cite{vanElp:1991}.}
\label{table:LCO_lat_param}
\end{table}
Regarding the band gap, the GGA value, 1.14~eV, is greatly improved by GGA+$U$ giving $2.90$~eV, in excellent agreement with the experimental measurement, 2.7 $\pm$ 0.3~eV~\cite{vanElp:1991}. This improvement is also consistent with a previous study on LiCoO$_2$~\cite{Miwa:2018} where similar values for the band gap were reported for both GGA and GGA+$U$. 

Similar to what we have done for CoO, we first present the high-frequency dielectric 
and Born effective charges tensors computed from DFPT and DFPT+$U$. Since the material is a polar insulator also without the Hubbard correction, a finite value of these quantities can be obtained also from GGA.
Table~\ref{table:LCO_epsilon_and_z} compares the two sets of results reporting the $xx$ and $zz$ components of both tensors (the $yy$ are equal to $xx$ by symmetry) that are diagonal in the rhombohedral representation. 
The comparison clearly shows that the Hubbard $U$ reduces the $xx$ component of the dielectric tensor  
quite significantly, while the variation of the $zz$ component is less pronounced. At the same time, the Born effective charges  are almost unaffected. 
Hence, differences in LO-TO splittings with and wihout $U$ correction are expected, as a consequence of 
different nonanalytic contributions to the dynamical matrix [see Eq.~\eqref{dyn_matrix_na}]. To the best of our knowledge, no experimental  $\vctr{Z}^*$ and $\boldsymbol{\epsilon}^{\infty}$ data  are available for this material.
\begin{table}[h]
\begin{center}
  \begin{tabular}{ccc}
    \hline\hline
                             &   \parbox{1.75cm}{DFPT}  &  \parbox{1.75cm}{DFPT+$U$}    \\ \hline
    $\epsilon^\infty_{xx}$   &   9.93  &         6.57                      \\ [1pt]
    $\epsilon^\infty_{zz}$   &   4.68  &         3.94                      \\ \hline
    $Z^*_{\mathrm{Co},xx}$   &   2.73  &         2.87                      \\ [1pt]
    $Z^*_{\mathrm{Co},zz}$   &   0.93  &         0.88                      \\ \hline
    $Z^*_{\mathrm{O},xx}$    &  -1.95  &        -2.01                      \\ [1pt]
    $Z^*_{\mathrm{O},zz}$    &  -1.32  &        -1.29                      \\ \hline
    $Z^*_{\mathrm{Li},xx}$   &   1.18  &         1.15                      \\ [1pt]
    $Z^*_{\mathrm{Li},zz}$   &   1.72  &         1.69                      \\ 
    \hline\hline
  \end{tabular}
\end{center}
\caption{LiCoO$_2$ high-frequency dielectric tensor and 
Born effective charges as computed using DFPT and DFPT+$U$. 
Both tensors are reported in the Cartesian framework and are diagonal, with equal $xx$ and $yy$ components.} 
\label{table:LCO_epsilon_and_z}
\end{table}  

\begin{figure}[h]
\begin{center}
   \includegraphics[width=0.48\textwidth]{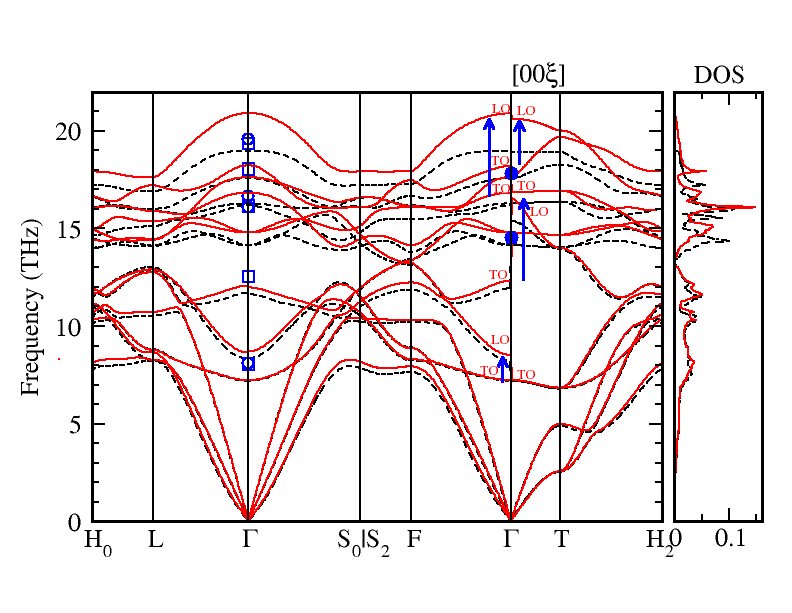}
   \caption{LiCoO$_2$ phonon dispersion in THz (left panel) and phonon density of states in states/THz/cell (right panel). DFPT (black dashed lines) and DFPT+$U$ (red solid lines). 
   Experimental data on Raman-active modes are  from    Ref.~\cite{Huang:1996} (filled circles). Infrared-active modes are from    Ref.~\cite{Huang:1996} (empty circles) and Refs.~\cite{Julien:2000,Julien:2002} (empty squares).  Other data from Refs.~\cite{CatroGarcia:2003,Inaba:1997, Zhecheva:1993, Ohzuku:1993} are redundant with the ones reported in the figure and are not shown for the sake of clarity. Arrows indicate the LO-TO splittings calculated with DFPT+$U$ and discussed in the text.}
\label{fig:LCO_disp}
\end{center}
\end{figure}

Fig.~\ref{fig:LCO_disp} shows the phonon dispersion of LiCoO$_2$ and compares the results of DFPT and DFPT+$U$ calculations with available Raman and infrared (IR) measurements at zone center.
The latter were performed on powder samples~\cite{Huang:1996, Julien:2000,Julien:2002, Inaba:1997, Zhecheva:1993, Ohzuku:1993, CatroGarcia:2003} (we are not aware of any INS 
experiment performed on single crystals of LiCoO$_2$ to sample its vibrational spectrum across the Brillouin zone)  and at finite temperature, 
while our simulations are done at 0\,K. 
Unlike in CoO and other TMOs~\cite{floris11}, here GGA results are already in acceptable agreement with the experiments. 
Interestingly, the Hubbard correction leaves the acoustic and lower optical phonon branches (up to about 13~THz) 
almost unchanged, while it affects the upper part of the 
spectrum more substantially. This subtle and
highly selective action of the $U$ on the phonon spectrum contrasts quite sharply with the significant effects it has on the electronic structure and, in particular, with the substantial widening of the band gap (see Table~\ref{table:LCO_lat_param}). The explanation of this selectivity resides in the fact that the low-frequency spectrum is dominated  by the  vibrations of highly mobile Li ions, not involving the stretching of the Co-O bonds. 
Above 13~THz instead, all modes imply the deformation of these bonds and are thus directly affected by the action of the Hubbard $U$ on Co 3$d$ states: the Raman active modes involving O vibrations (14.8~THz, 17.6~THz), the O-Co LO and TO modes (16.4--16.7~THz) and the topmost LO modes (20.6--20.9~THz).

Concerning the comparison with the experiments, we note an excellent agreement with the Raman active modes (Fig.~\ref{fig:LCO_disp}, filled circles at $\Gamma$), slightly improved by DFPT+$U$. The computed Raman frequencies, (14.80 and 17.63~THz) are also in very good agreement with those obtained in Ref.~\cite{Miwa:2018} (14.72 and 17.60~THz, for the E$_g$ and A$_{1g}$ modes, respectively) which provides a further validation of our implementation. Raman-active modes are continuous  along F--$\Gamma$--T since they do not excite  fluctuating dipoles  and hence there is no macroscopic electric field associated with them. IR active modes, instead, experience LO-TO splittings that cannot be resolved by experiments on powder samples. This fact makes the direct comparison with the available experimental data  more problematic (one should perhaps understand experimental points as average frequencies of split LO-TO couples along all the possible directions to $\Gamma$). While it is difficult to assess the quality of DFPT and DFPT+$U$ results for IR modes, it is evident that DFPT+$U$ frequencies are increasingly blue-shifted compared to GGA  in the upper region of the spectrum, with a particularly large difference for the topmost mode. This result is probably related to the widening of the band gap produced by the Hubbard correction, which is  also reflected in the strongly reduced value of the $\epsilon^\infty_{xx}$ entry of the dielectric tensor compared to GGA (see Table~\ref{table:LCO_epsilon_and_z}), implying larger LO-TO splittings [see Eq.~\eqref{dyn_matrix_na}]. 

It is worth it at this point to discuss in some detail the nature of these splittings. This is particularly important in view of future INS experiments on LiCoO$_2$ single crystals, which we hope to stimulate with our calculations. 
Referring to the DFPT+$U$ spectrum and focusing for simplicity on the F--$\Gamma$-T panels, the splittings are highlighted in Fig.~\ref{fig:LCO_disp} by blue arrows. 
The first one involves 
the doublet of TO modes (along $\Gamma$--T) at 7.2\,THz that consist mostly of Li vibrations. One of these modes maintains its transverse character also along F--$\Gamma$ and is continuous; the other acquires a LO component that shifts up its energy by 1.3 THz, due to a  coupling with the macroscopic electric field. 
The TO mode along F--$\Gamma$ (12.3\,THz) moves Li ions almost parallel to the [001] direction;  
therefore, it almost coincides with the LO mode  along $\Gamma$--T, 
whose energy has shifted up by about 4.24\,THz. 
The third TO $\Gamma$-T doublet at 16.8\,THz moves the Co and O sublattices in counter-phase.
In analogy with the first doublet discussed above, only one of these modes remains TO along F--$\Gamma$, while the other acquires a longitudinal component that shifts its energy up by almost 4\,THz, becoming the  highest frequency mode along the latter direction. At the same time, the
mostly TO  mode along F-$\Gamma$ (second highest energy at 18.2\,Thz)  becomes the LO highest energy branch along $\Gamma$--T.

The same analysis  
is qualitatively valid also for the GGA phonon dispersion  
even if, as anticipated above, the entity of the splittings and discontinuities differ.

It is interesting to note that the 
nature of the discontinuities (e.g. in the topmost LO branch) is different in CoO and LiCoO$_2$: in the first it is related to the magnetic order and the rhombohedral distortion, in the second it descends from the different coupling of different vibrations with the electric fields (in fact F--$\Gamma$ and $\Gamma$-T are not crystallographically equivalent). 

We close this section by mentioning  
previous computational studies on the 
vibrational properties of LiCoO$_2$. 
In Refs.~\cite{Gong:2013, Du:2016}, the authors computed the phonon dispersion 
using the Hubbard correction; their calculations were based, however, on a frozen-phonon approach and did not include LO-TO splittings, 
that are instead a key feature of the dispersion.
In Ref.~\cite{yang19} 
the vibrational spectra of LiCoO$_2$ were used as a starting point to analyze its anharmonic lattice dynamics and 
heat transport. This work 
also compared the spectra obtained from LDA, LDA+$U$ and hybrid HSE06 functionals.
In Ref.~\cite{Miwa:2018}, the generalized DFPT approach was also used 
with a Hubbard correction including $U$ and $J$.
However, only Raman-active modes were presented and no phonon dispersion was shown.

\section{Conclusions}
\label{concl}
 
We presented, within the ultrasoft pseudopotentials (US PPs) formalism, a comprehensive theoretical derivation to include the DFT+$U$ functional in the framework of the density functional perturbation theory (DFPT), used for the calculation of phonons and related quantities.  
The approach, denoted DFPT+$U$, represents a powerful  tool for accurate and efficient linear-response calculations in strongly correlated materials. This formulation develops an important extension to US PPs of a previous implementation of DFPT+$U$ deviced for NC PPs~\cite{floris11}. A similar generalization was developed independently in Ref.~\cite{Miwa:2018}. Starting from the DFT+$U$ expressions of total energy and  forces, we show how Hubbard-related DFPT terms appear in the perturbed Hubbard potential, 
in the matrix of force constants, and in the electronic dielectric tensor and Born effective charges that determine the nonanalytical part of the dynamical matrix of polar insulators. The  terms 
are classified on the basis of their  self-consistent or bare  nature (depending on whether or not they require the evaluation of the Kohn-Sham wave functions response) and of their specificity to the US PPs formalism. 
This is useful to reproduce the implementation in different codes and to stimulate future developments using linear-response theory on a DFT+$U$ ground state.
The formalism is also extended to metallic systems, where a smeared Fermi distribution function is necessary. This paves the way to understand, for example, the possible role of correlation in describing phonon anomalies or the electron-phonon coupling in doped insulators exhibiting a metallic behavior.

DFPT+$U$ is applied here to  study  the vibrational spectra of two insulating Co oxides, CoO and LiCoO$_2$.
In CoO the Hubbard correction leads to a dramatic, qualitative improvement over the GGA results, eliminating the dynamical instabilities associated to the spurious metallic character predicted by the noncorrected functional. 
Through the stabilization of an insulating ground state, the Hubbard correction also refines quantitatively the
structural and vibrational properties of the material, achieving an  
excellent agreement with the experiments. 

For LiCoO$_2$, the Hubbard correction improves substantially 
the agreement with experimental results for the band gap and the equilibrium lattice parameters. Regarding the vibrational spectrum, its effect is more subtle and energy-dependent. While, importantly,  it has marginal effects on the acoustic and lower optical modes, where GGA already performs well, it slightly improves 
Raman-active frequencies. 
The comparison with IR-active modes measured from powder samples is instead more problematic.  
In general the Hubbard-corrected IR-active modes are blue-shifted in the upper part of the spectrum, where vibrations involve more significantly Co ions. This result is consistent with a better localization of  Co 3$d$ 
electrons, that widens the band gap and attenuates the electronic screening, but would require a comparison with INS experiments on crystalline samples to be precisely assessed. 

Finally,
we stress that our approach, at variance with other works, is fully {\it ab initio}, with no input from experiments (e.g., on $\epsilon^\infty$ or $Z^*$) nor adjustable parameters (e.g., the Hubbard $U$). The quantitative agreement of our results with experiments is thus  
a further proof of the
effectiveness of DFPT+$U$ and the self-consistent evaluation of the Hubbard interaction parameters.

\section*{Acknowledgement}
This work was supported by the Deutsche Forschungsgemeinschaft. 
This research was partially supported by the Swiss National Science Foundation (SNSF), through Grant No. 200021-179138, and its National Centre of Competence in Research (NCCR) MARVEL. Computer time was provided by CSCS (Piz Daint) through project No.~s836, and by CINECA through an award under the ISCRA initiative.


\begin{appendices}

\section*{Appendix: DFPT+$U$ for metals}
\label{metals}

In this Appendix we extend the DFPT+$U$ formalism with US PPs to metallic systems (see also Refs.~\cite{degironcoli95,dalcorso01,baroni01}).
This extension improves the treatment of metals whenever the localization of some of their valence electrons plays a relevant role, e.g. in determining fine details of the Fermi surface, phonon dispersions and electron-phonon couplings. Calculations for metallic systems and magnetic insulators~\cite{Note_tech} require a smearing of the distribution function around the Fermi level, useful to avoid numerical instabilities. 
We indicate by $\thetFis \equiv \tilde{\theta}\left[(\varepsilon_F - \eis)/\eta\right]$ the occupation of the KS state ($i \sigma$). Here $\tilde{\theta}$ represents a smooth generalization of the Fermi-Dirac function centered at the Fermi energy $\varepsilon_F$, whose shape is controlled by the specific definition adopted and by the smearing width $\eta$ (see Refs.~\cite{methfessel89, marzari99} for notable examples). States within $\eta$ from the Fermi level assume  fractional occupations.

In metals, the occupation matrices $n^{I\sigma}_{m_1m_2}$, Eq.~\eqref{ns}, are generalized in a similar way to the
charge density to account for the fractional occupation of KS states: 
\be
n^{I\sigma}_{m_1m_2} = \sumi \thetFis \bra{\psis} \Proj \ket{\psis} \,.
\label{ns_met}
\ee
Hence, their SCF derivative [Eq.~\eqref{dns}] contains the extra contribution stemming from the variation of the KS occupations around the Fermi level:
\be
\Delta^{\mu}_{\mathrm{met}} n^{I\sigma}_{m_1m_2} =
\sumi \frac{d\thetFis}{d\mu} \bra{\psis} \Proj \ket{\psis} .
\label{dns_met}
\ee
This  can be computed as~\cite{dalcorso01}: 
\bea
 \frac{d\thetFis}{d\mu} = 
\frac{1}{\eta} \, \tilde{\delta}_{F,i\sigma} 
\left[ \frac{d\varepsilon_F}{d\mu} - \frac{d\eis}{d\mu} \right] \,,
\label{dtheta_var}
\eea
where $\tilde{\delta}_{F,i\sigma} \equiv \tilde{\delta}[(\varepsilon_F-\eis)/\eta]$ approximates the Dirac's $\delta$ function in the limit of vanishing $\eta$. As for the derivatives $\frac{d\varepsilon_F}{d\mu}$ and  $\frac{d\eis}{d\mu}$ [Eq.~\eqref{deis2}], we refer the reader also to the discussion in Sec.~II.C.4 of Ref.~\cite{baroni01}. Since these 
derivatives depend on the response of the KS and Hubbard potentials, Eq.~\eqref{dns_met} must be computed at every iteration during the solution of the Sternheimer equation.
Other terms of the response occupation matrix, namely $\frac{\de n^{I\sigma}_{m_1m_2} }{\de \lambda}$, $\Dn$ and $\delta^\mu n^{I\sigma}_{m_1m_2}$ [Eqs.~\eqref{dnbare}, \eqref{dns_cond} and \eqref{dnsv}], also require reconsideration in the case of metals. The bare term simply becomes:
\be
\frac{\de n^{I\sigma}_{m_1m_2} }{\de \lambda} = \sumi \thetFis  
\Bra{\psis} \frac{\de \Proj}{\de \lambda} \Ket{\psis} \,.
\label{dnbare_met}
\ee
In order to generalize the other two terms (SCF and US PPs-specific), one has to extend the Sternheimer equation~\eqref{dks2} to fractional occupations. 
This is done in such a way that the  Sternheimer equation for metals  can still be formally written in the insulator-like form of Eq.~\eqref{dks2}~\cite{degironcoli95}. 
The mathematical details
can be found in Ref.~\cite{dalcorso01} and will not be addressed here. We will focus, instead, on the generalization of the quantities pertaining to the Hubbard functional.  The generalized solution of the Sternheimer equation~\eqref{dks2} can be formally written as (cf. with the second term in Eq.~(26) in Ref.~\cite{dalcorso01}):
\begin{eqnarray}
\ket{\Dpsis} & = & 
\sumj \frac{\thetFis - \thetFjs}{\eis - \ejs} \thetjsis \ket{\psjs} \nn \\
& & \times \Bra{\psjs} \left[ \frac{d\hat{V}^\sigma_\mathrm{KS}}{d\mu} + 
\frac{d\hat{V}^\sigma_\mathrm{Hub}}{d\mu} - \eis \frac{\de \hat{S}}{\de \mu} \right] \Ket{\psis} \,, \nn \\
& & 
\label{proc_met}
\end{eqnarray}
where $\theta_{j\sigma,i\sigma} \equiv \theta\left[(\ejs - \eis)/\eta\right]$, and  
$\theta(\varepsilon)=\mathrm{erfc}(-\varepsilon)/2$.
Equation~\eqref{proc_met} is the metallic counterpart of Eq.~\eqref{projc} for insulators. Based on this expression and using Eq.~\eqref{dns_met}, it can be shown that the SCF term of the response occupation matrix~\eqref{dns_cond} is generalized as:
\begin{eqnarray}
\Dnmet & \equiv & 
\sumi \left[ \bra{\Dpsismet} \Proj \ket{\psis} \right. \nn \\
& & \hspace{0.8cm} + \, \left. \bra{\psis} \Proj \ket{\Dpsismet} \, \right] \,,
\label{dns_cond_met}
\end{eqnarray}
where we  defined~\cite{dalcorso01}:
\be
\ket{\Dpsismet} = \ket{\Dpsis} +
\frac{1}{2\eta} \, \tilde{\delta}_{F,i\sigma} \frac{d\varepsilon_F}{d\mu} \, \ket{\psis} \,.
\label{dpsi_met}
\ee
Note that in the derivation of Eq.~\eqref{dns_cond_met} we   used the fact that the $j=i$ term (which is singular) in Eq.~\eqref{proc_met} corresponds to the second term in Eq.~\eqref{dtheta_var} multiplied by $\frac{1}{2}\ket{\psis}$. 
Finally, the remaining term $\delta^\mu n^{I\sigma}_{m_1m_2}$ can still be 
written as in Eq.~\eqref{dnsv}, as long as $\ket{\delta^{\mu} \psis}$ is generalized to account for the smearing of the distribution function~\cite{dalcorso01}:
\begin{eqnarray}
\ket{ \delta^{\mu} \psis } & = & 
\sumj \Ket{\psjs} \Bra{\psjs} \frac{\de \hat{S}}{\de\mu} \Ket{\psis} \nn \\
& & \hspace{0.5cm} \times \left[ \thetFis \thetisjs + \thetFjs \thetjsis \right] \,.
\label{dpsiorth_met}
\end{eqnarray}
To summarize, the total response occupation matrix in the case of metals reads:
\begin{eqnarray}
\frac{d n^{I\sigma}_{m_1m_2}}{d\mu} & = & 
\frac{\de n^{I\sigma}_{m_1m_2} }{\de \mu} + \Dnmet + \delta^{\mu} n^{I\sigma}_{m_1 m_2} \,,
\label{dns_scf_met}
\end{eqnarray}
where $\frac{\de n^{I\sigma}_{m_1m_2} }{\de \mu}$ is given by Eq.~\eqref{dnbare_met}, $\Dnmet$  by Eqs.~\eqref{dns_cond_met} and \eqref{dpsi_met}, and $\delta^{\mu} n^{I\sigma}_{m_1 m_2}$ by Eqs.~\eqref{dnsv} and \eqref{dpsiorth_met}.

Finally, let us consider the matrix of interatomic force constants. In the metallic case,  the force $F_\lambda$ [Eq.~\eqref{for2}] 
has the prefactor $\thetFis$ in the summation over $i$ and $\sigma$. Consequently, the matrix of force constants $C_{\mu\lambda}$ [Eq.~\eqref{dyndt}]  has an extra term that accounts for the variation of the occupation of KS states [Eq.~\eqref{dtheta_var}]:
\begin{eqnarray}
C_{\mu\lambda}^\mathrm{met} & = &
\sumis \frac{d\thetFis}{d\mu} \, 
\Bra{\psis} \left[ \frac{\de \hat{V}_\mathrm{KS}^{\sigma}}{\de \lambda} \right. \nn \\ 
& & \hspace{1.3cm} + \, \left. \vtilde \frac{\de \Prj}{\de \lambda} - 
\eis \frac{\de \hat{S}}{\de\lambda} \right] \Ket{\psis}.
\label{force_const_met}
\end{eqnarray} 
Equations~\eqref{force_const_a}, \eqref{deis1}, and \eqref{deis_hub} have also a prefactor $\thetFis$ in the summation over $i$ and $\sigma$.
Using Eqs.~\eqref{deis2}, \eqref{dtheta_var} and \eqref{force_const_met},
it can be shown that the Hubbard terms~\eqref{force_const_b},
\eqref{force_const_c}, and \eqref{force_const_d}
can still be written in the insulator-like form illustrated in Sec.~\ref{sec:Hub_force_constants}, but with 
$\ket{\delta^{\mu} \psis}$ being generalized as in Eq.~\eqref{dpsiorth_met},
and $\ket{\Dpsis}$ [in Eq.~\eqref{force_const_b}] being replaced by
$\ket{\Dpsismet}$, defined in Eqs.~\eqref{proc_met} and \eqref{dpsi_met}.

The regrouping of the Hubbard terms in the matrix of force constants
proposed at the end of Sec.~\ref{sec:Hub_force_constants} is still valid in
the metallic case. 
The difference is that now $\frac{\de n^{I\sigma}_{m_1m_2} }{\de \mu}$
is defined by Eq.~\eqref{dnbare_met} (similar statement holds for the second
bare derivative of the occupation matrix),
$\delta^{\mu} n^{I\sigma}_{m_1 m_2}$  by 
Eqs.~\eqref{dnsv} with $\delta^{\mu} \psis$ defined as in Eq.~\eqref{dpsiorth_met},
and, finally, $\ket{\Dpsis}$ and $\Dn$ in Eq.~\eqref{force_const_3} are replaced by
their metallic counterparts $\ket{\Dpsismet}$ and $\Dnmet$, defined in Eqs.~\eqref{dpsi_met}
and \eqref{dns_cond_met}, respectively.

The implementation of DFPT+$U$ for metallic systems, presented in this Appendix, is validated in Sec.~\ref{secSM:metals} of the SM~\cite{SupplementalMaterial} by comparing its results with those from finite differences.

\end{appendices}


\begin{thebibliography}{116}
\bibitem [{Not({\natexlab{a}})}]{Note_tech}%
  \BibitemOpen
  \href@noop {} {} \bibinfo {note} {{E}ven for insulators, a
  finite smearing might be needed technically to avoid that codes impose double
  occupancy of single-particle states, thus stabilizing nonmagnetic ground
  states. This is particularly important for magnetic insulators as, e.g.,
  transition-metal oxides.}\BibitemShut {Stop}%
\bibitem [{Not({\natexlab{b}})}]{Note_ortho}%
  \BibitemOpen
  \href@noop {} {} \bibinfo {note} {{B}ased on
  Eqs.~\eqref{ortho}, ~\eqref{ks} and ~\eqref{dpsi2}, we can derive the general
  expression of the forces: \newline \scriptsize{$F_{\lambda} = -\frac{d
  E_{\mathrm{DFT}+U}}{d \lambda} = -\sum_{i\sigma} \int \frac{\delta
  E_{\mathrm{DFT}+U}}{\delta \psis^*({\mathbf r})}\frac{d\psis^*({\mathbf
  r})}{d\lambda}d{\mathbf r} - \int \frac{\delta E_{\mathrm{DFT}+U}}{\delta
  \psis({\mathbf r})}\frac{d\psis({\mathbf r})}{d\lambda}d{\mathbf r} -
  \frac{\partial E_{\mathrm{DFT}+U}}{\partial \lambda}=$}
  \scriptsize{$-\sum_{i\sigma} \left\{ \Bra{\frac{d \psis}{d \lambda}}
  \hat{H}^\sigma \Ket{\psis} + \Bra{\psis} \hat{H}^\sigma \Ket{\frac{d \psis}{d
  \lambda}} + \Bra{\psis} \frac{\partial \hat{H}^\sigma}{\partial \lambda}
  \Ket{\psis} \right\} =$} \scriptsize{$- \sum_{i\sigma} \left\{
  \eis\left[\Bra{\frac{d\psis}{d\lambda}} \hat{S} \Ket{\psis} + \Bra{\psis}
  \hat{S} \Ket{\frac{d\psis}{d\lambda}} \right] + \Bra{\psis} \frac{\partial
  \hat{H}^\sigma}{\partial \lambda} \Ket{\psis} \right\} =$}
  \scriptsize{$\sum_{i\sigma} \left\{ \eis \Bra{\psis} \frac{\de
  \hat{S}}{\de\lambda} \Ket{\psis} - \Bra{\psis} \frac{\partial
  \hat{H}^\sigma}{\partial \lambda} \Ket{\psis} \right\}.$} \newline
  \normalsize $\hat{H}^\sigma$ is defined in
  Eqs.~\eqref{eq:Hamiltonian}-\eqref{vtilde}, and by virtue of Eq.~\eqref{pot},
  for DFT+$U$ calculations the last term of the previous equation can be
  expressed as $\Bra{\psis} \frac{\partial \hat{V}_{KS}}{\partial \lambda}
  \Ket{\psis}+\frac{\partial E_U}{\partial \lambda}$, hence Eq.~\eqref{for}.
  For more details, the reader is referred to
  Ref.~\cite{dalcorso01}.}\BibitemShut {Stop}%
\bibitem [{Not({\natexlab{c}})}]{Note_Ewald}%
  \BibitemOpen
  \href@noop {} {} \bibinfo {note} {{T}he terms coming from
  Ewald sums and non-linear core corrections have been omitted from
  Eq.~\eqref{for}. An explicit account of these terms in the context of US PPs
  can be found in Ref.~\cite{dalcorso01}.}\BibitemShut {Stop}%
\bibitem [{Not({\natexlab{d}})}]{Note_force_cancellation}%
  \BibitemOpen
  \href@noop {} {} \bibinfo {note} {{N}ote that the
  $\frac{\de V^{\sigma}_\mathrm{Hxc}(\vr)}{\de \lambda}$ contributions to the
  bare derivative of $V^{\sigma}_\mathrm{KS}$ cancel out, and only $\frac {\de
  V_\mathrm{loc}(\vr)}{\de \lambda}$ survives~\cite{dalcorso01}.}\BibitemShut
  {Stop}%
\bibitem [{Not({\natexlab{e}})}]{Note_LHS_Sternheimer}%
  \BibitemOpen
  \href@noop {} {} \bibinfo {note} {{W}hen the system in
  Eq.~\eqref{dks2} is solved by non-iterative methods, a term $\alpha
  \hat{\mathcal{O}}^{\sigma}$, with $\hat{\mathcal{O}}^{\sigma} =
  \sum_j^\mathrm{occ} \hat{S} |\psjs\rangle\langle\psjs| \hat{S}$ and $\alpha =
  2 \, (\mathrm{max}[\eis] - \mathrm{min}[\eis])$ is typically added to the
  left-hand side of Eq.~\eqref{dks2} in order to avoid possible singularities
  of the operator \cite{baroni01}}\BibitemShut {NoStop}%
\bibitem [{Not({\natexlab{g}})}]{Note_Ud}%
  \BibitemOpen
  \href@noop {} {} \bibinfo {note} {{The Hubbard $U$
  parameter can be thought as an average screened interaction among atomic
  orbitals. It thus depends on atomic positions both explicitly (through the
  center of atomic orbitals) and implicitly (through the readjustment of KS
  states contributing to its screening). A numerical method to evaluate the
  first order derivative of $U$ with respect to the atomic positions has been
  proposed in Ref.~\cite{kulik11}, based on the linear-response approach to
  compute $U$ discussed in Ref.~\cite{cococcioni05}. However, to the best of
  our knowledge, no attempt has been made to evaluate its second order
  derivatives.}}\BibitemShut {Stop}%
\end{thebibliography}

%


\clearpage
\clearpage 
\setcounter{page}{1}
\renewcommand{\thetable}{S\arabic{table}}  
\setcounter{table}{0}
\renewcommand{\thefigure}{S\arabic{figure}}
\setcounter{figure}{0}
\renewcommand{\thesection}{S\arabic{section}}
\setcounter{section}{0}
\renewcommand{\theequation}{S\arabic{equation}}
\setcounter{equation}{0}
\onecolumngrid

\begin{center}
\textbf{\large Supplemental Material for} 
\vskip 0.2 cm
\textbf{\large Hubbard-corrected density functional perturbation theory with ultrasoft pseudopotentials}
\vskip 0.2cm
A. Floris, I. Timrov, B. Himmetoglu, N. Marzari, S. de Gironcoli, and M. Cococcioni
\end{center}

\section{Benchmark of DFPT+$U$ for US and PAW PPs on polar insulators}
\label{secSM:insulators}

\begin{figure*}[h!]
\begin{center}
  \subfigure[]{\includegraphics[width=0.49\textwidth]{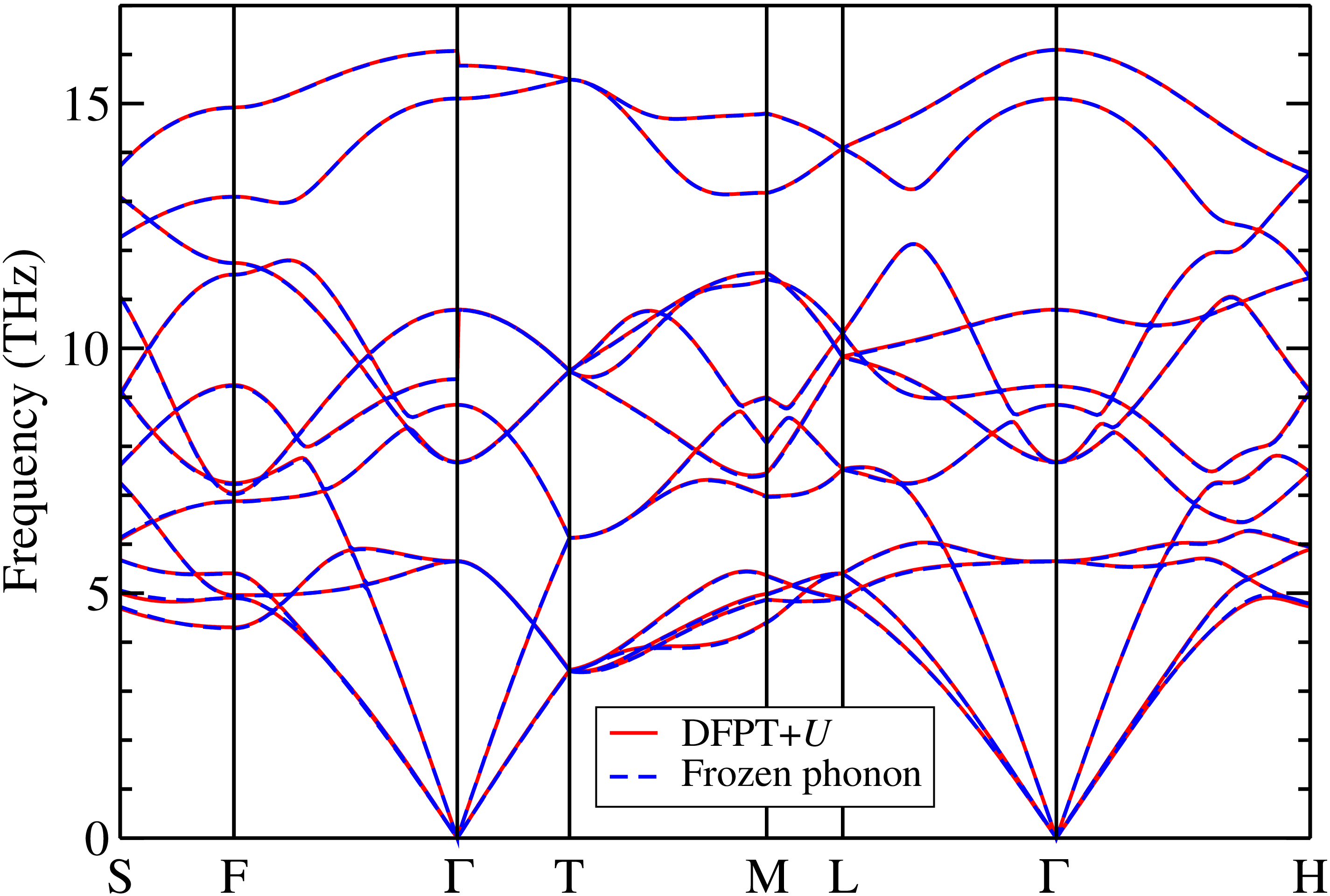}\hspace{0.05cm}}
  \subfigure[]{\includegraphics[width=0.49\textwidth]{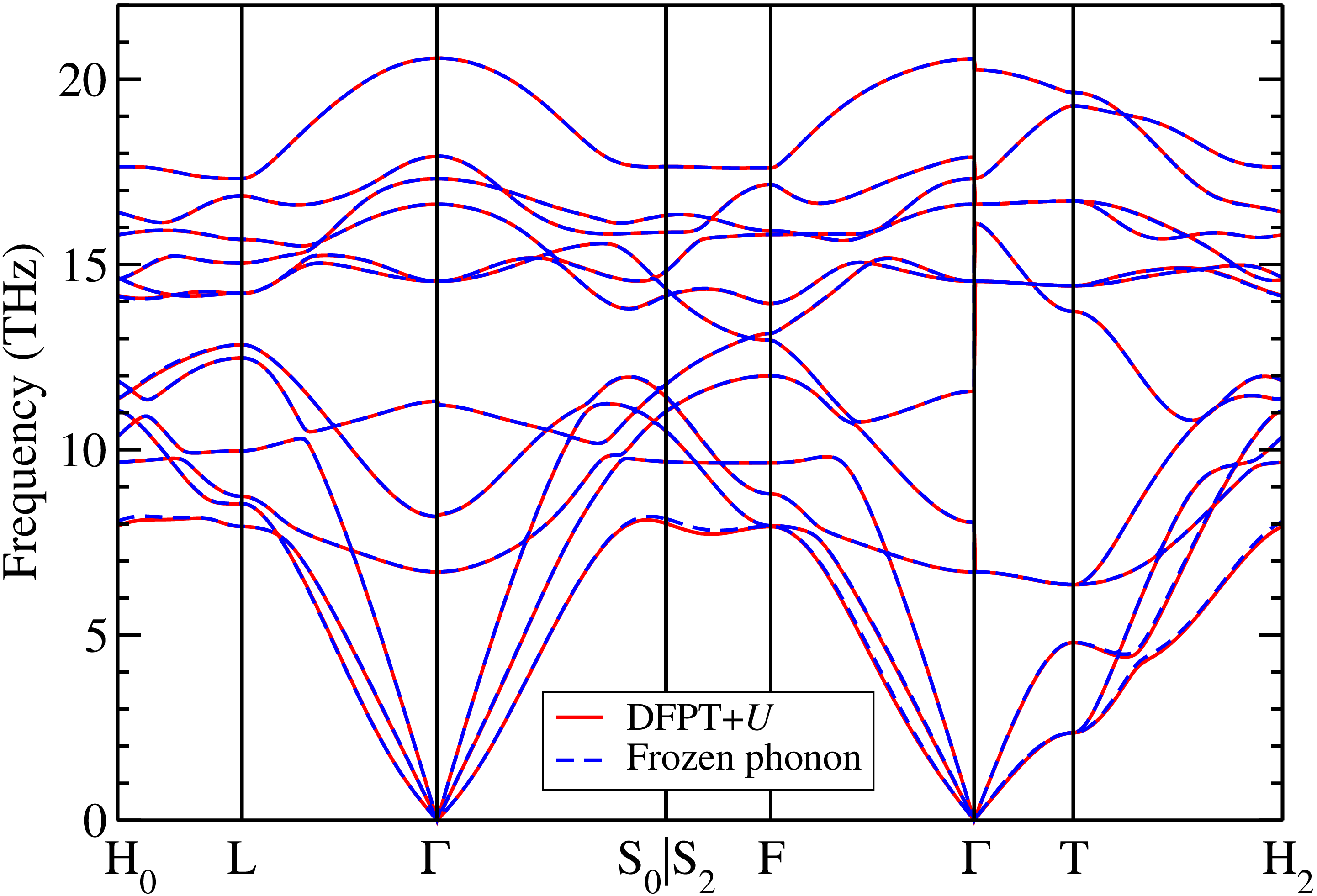}}
  \caption{Comparison of the phonon dispersions computed using DFPT+$U$ and the frozen-phonon approach 
  for (a)~CoO using US PPs, and (b)~LiCoO$_2$ using PAW PPs. The nonanalytic term of the dynamical matrix (as obtained from DFPT+$U$) was included in all cases (see Sec.~\ref{secSM:nonanalytic_term} for more details).}
  \label{figSM:disp}
\end{center}
\end{figure*}

{\bf Discussion.} The aim of this section is to 
benchmark our DFPT+$U$ implementation for both ultrasoft (US) and projector-augmented wave (PAW) pseudopotentials (PPs). 
This task is realized by contrasting the phonon dispersion obtained from DFPT+$U$ against the one from frozen-phonon calculations. In this section we focus specifically on the study of polar insulators, while  metallic systems will be tested separately, in Sec.~\ref{secSM:metals}, so to ascertain the validity of all the terms related to the derivatives of the fractional occupations of KS states.
Test cases here 
are CoO and LiCoO$_2$, whose DFPT+$U$ results are discussed in the main text.
Figure~\ref{figSM:disp} shows a comparison of the phonon dispersions computed using DFPT+$U$ and the frozen-phonon approach for CoO using US PPs and for LiCoO$_2$ using PAW PPs. The two approaches are in excellent agreement and the phonon dispersions are 
superimposed
for both PPs types. These results provide a solid validation of our DFPT+$U$ implementation and an indirect test also of the correctness of DFT+$U$ forces.

\vskip 0.2cm

{\bf Technical details.} The technical details of the calculations  are the same as described in Sec.~V of the main text, apart from few differences discussed in the following. For CoO we used US PPs~[1] and for LiCoO$_2$ we used PAW PPs~[2]. For both systems the DFPT+$U$ calculations were performed using the uniform $\Gamma$-centered $8 \times 8 \times 8$ and $4 \times 4 \times 4$ $\mathbf{k}$ and $\mathbf{q}$ point meshes, respectively. Consistently, the frozen-phonon calculations were run with supercells of size $4 \times 4 \times 4$ and a $\mathbf{k}$ point mesh of size $2 \times 2 \times 2$.
Frozen-phonon pre- and post-processing analysis were performed using the Phonopy code~[3], based on DFT+$U$ forces computed with   the \textsc{Quantum-ESPRESSO} package~[4,5] that was also employed for  the DFPT+$U$ calculations. For consistency, for the $3d$ states of Co in CoO
we have used the \textit{ab initio} self-consistent Hubbard $U$ parameter (4.55~eV) and the same crystal geometry  presented in the main text (see Table~I).  
For LiCoO$_2$, instead, we re-computed the Hubbard $U$ parameter and re-optimized the crystal geometry self-consistently using the PAW PPs and obtained $U = 7.01$~eV.

\vskip 0.3cm

\newpage
\section{Benchmark of the nonanalytic term of the dynamical matrix in DFPT+$U$} 
\label{secSM:nonanalytic_term}

{\bf Discussion.} 
The aim of this section is to benchmark 
the Hubbard contribution to the Born effective charges 
$\vctr{Z}^*$ and to the high-frequency (electronic) dielectric tensor $\boldsymbol{\epsilon}^{\infty}$ that, as explained in Sec.~IV.C of the main text, are used to evaluate the nonanalytic (NA) term of the dynamical matrix of polar insulators at zone center.
Correcting the frequency of longitudinal optical modes through re-establishing the coupling with the macroscopic electric field they generate, this term makes the phonon energies at $\Gamma$ consistent with those of $\vctr{q}$-points in its proximity ( $\vctr{q}\rightarrow \mathbf{0}$  
- where the coupling with the electric field is automatically included) and allows to capture the LO-TO splittings in the phonon dispersion of polar insulators at zone center~[6,7]. 
Consequently, the test of the Hubbard contribution to the nonanalytic part of the dynamical matrix is an indirect one and is based on comparing the phonon dispersion  
in the long-wavelength limit,
after nonanalytic terms are included in the dynamical matrix, with the vibrational frequencies computed explicitly at $\vctr{q}$-points in its proximity along various directions.
Since LO-TO splittings only affect the phonon dispersion of polar insulators we chose again CoO and LiCoO$_2$ as test cases that were treated with US and PAW PPs, respectively.

\begin{figure*}[h!]
\begin{center}
  \subfigure[]{\includegraphics[width=0.485\textwidth]{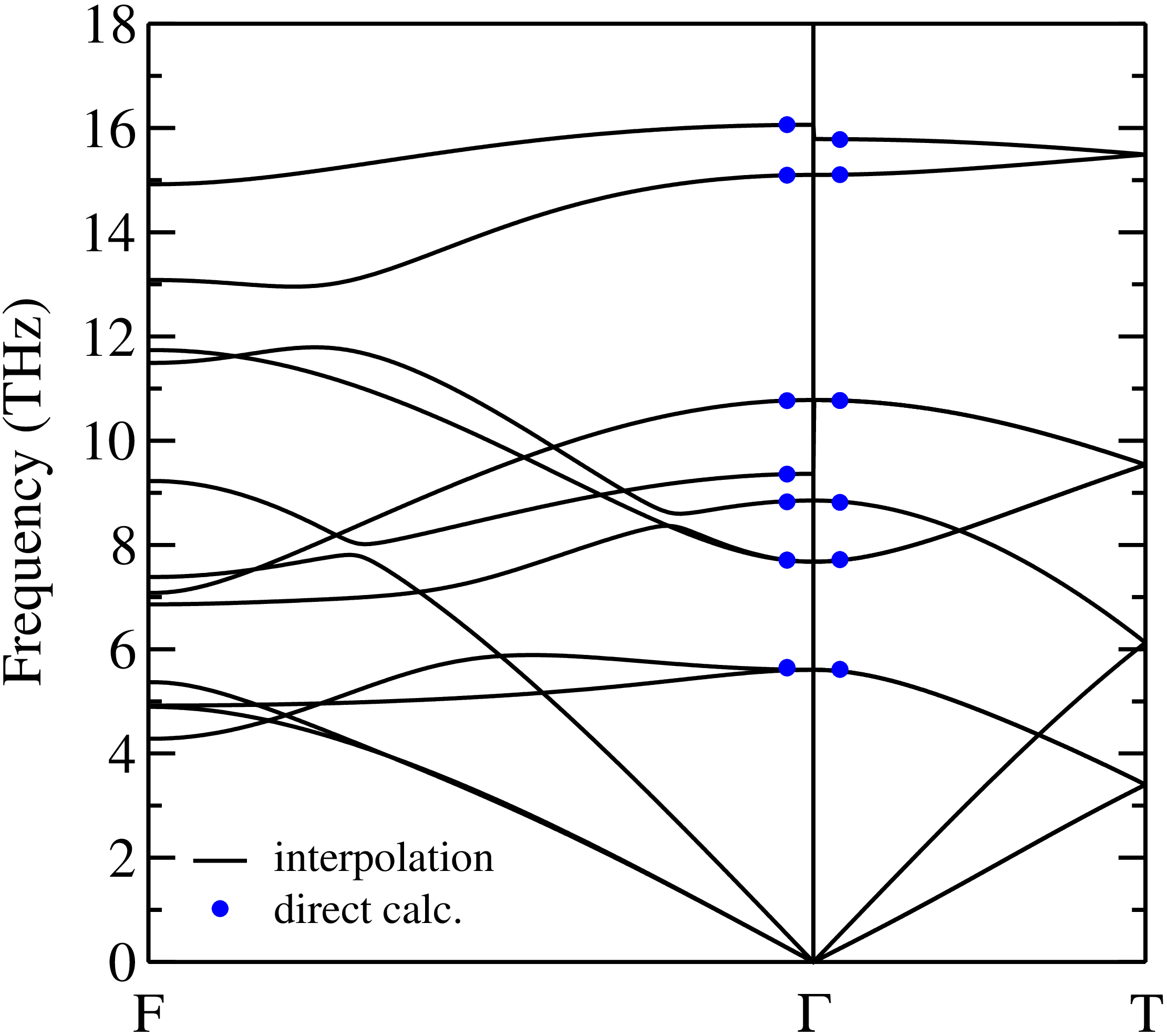}\hspace{0.05cm}}
  \subfigure[]{\includegraphics[width=0.485\textwidth]{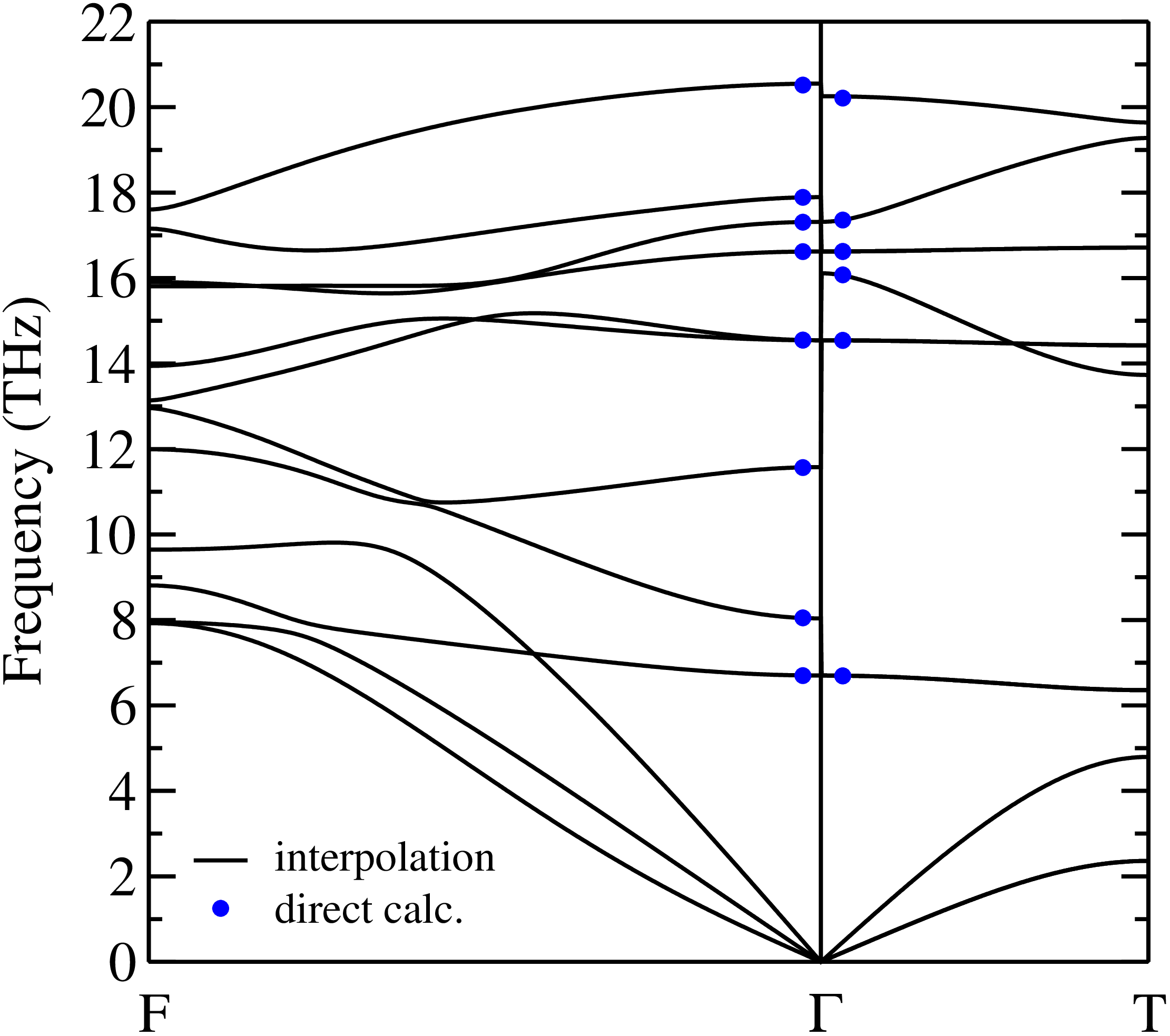}}
  \caption{DFPT+$U$ phonon dispersions of CoO (a) and LiCoO$_2$ (b) calculated with US and PAW PPs, respectively. Black solid lines are obtained by Fourier-interpolating the results of direct calculations on a $4 \times 4 \times 4$ $\mathbf{q}$ grid with nonanalytic terms at $\vctr{q}=\mathbf{0}$; blue dots indicate the frequencies of optical modes directly calculated at $\mathbf{q}$ points in the close proximity of $\Gamma$. 
  \label{figSM:disp_nonanalytic_term}}
\end{center}
\end{figure*}

Fig.~\ref{figSM:disp_nonanalytic_term} shows the result of this benchmark. Blue dots indicate the frequencies of phonons explicitly calculated at 
$\vctr{q}$-points located in the close proximity of $\Gamma$. Black solid lines represent the phonon dispersions obtained 
from correcting the dynamical matrix at $\Gamma$ with direction-specific nonanalytic terms. 
A remarkable agreement can be observed between black lines and blue dots for both US (CoO) and PAW PPs (LiCoO$_2$), 
The excellent agreement
we obtain when NA terms are included 
validate  
the Hubbard contributions to 
$\vctr{Z}^*$ and $\boldsymbol{\epsilon}^{\infty}$. 

\vskip 0.2cm

{\bf Technical details.} The technical details of the calculations are the same as in Sec.~\ref{secSM:insulators}. 
CoO direct calculations  
were performed at $\mathbf{q} = \frac{2\pi}{a}(0.02, -0.02, -0.02)$ along $\Gamma$-F and at $\mathbf{q} = \frac{2\pi}{a}(0.02, 0.02, 0.02)$ along $\Gamma$-T, 
with $a = 8.00$~Bohr (see blue dots in Fig. \ref{figSM:disp_nonanalytic_term}(a)).  For LiCoO$_2$, 
at $\mathbf{q} = \frac{2\pi}{a}(0.00, -0.03, 0.01)$ along $\Gamma$-F and at $\mathbf{q} = \frac{2\pi}{a}(0.00, 0.00, 0.035)$ along $\Gamma$-T, 
with $a = 9.37$~Bohr (see blue dots in Fig. \ref{figSM:disp_nonanalytic_term}(b)).

\vskip 0.2cm

\newpage
\section{Benchmark of DFPT+$U$ for metallic systems}
\label{secSM:metals}

\begin{figure*}[h!]
\begin{center}
  \subfigure[]{\includegraphics[width=0.49\textwidth]{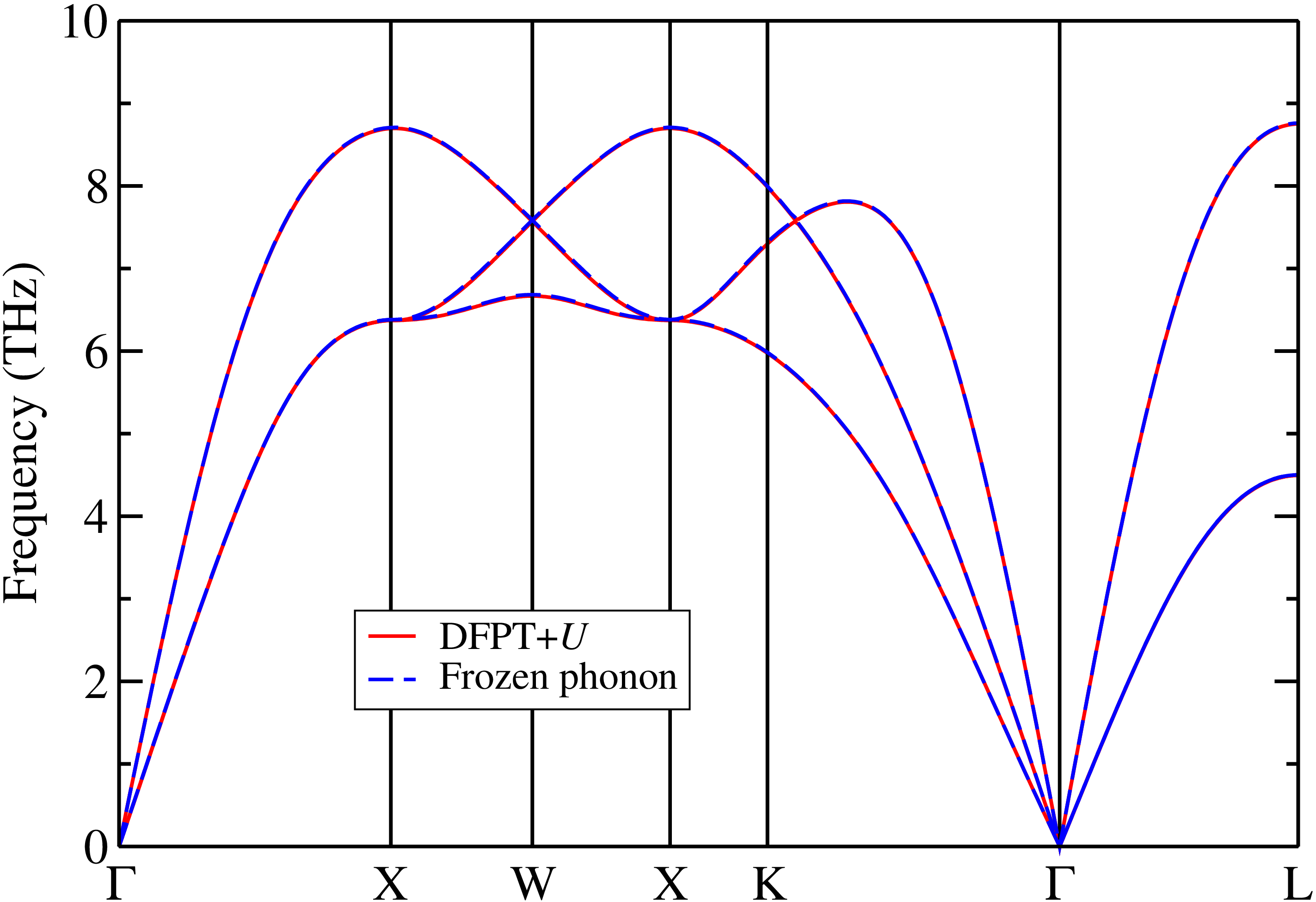}\hspace{0.05cm}}
  \subfigure[]{\includegraphics[width=0.49\textwidth]{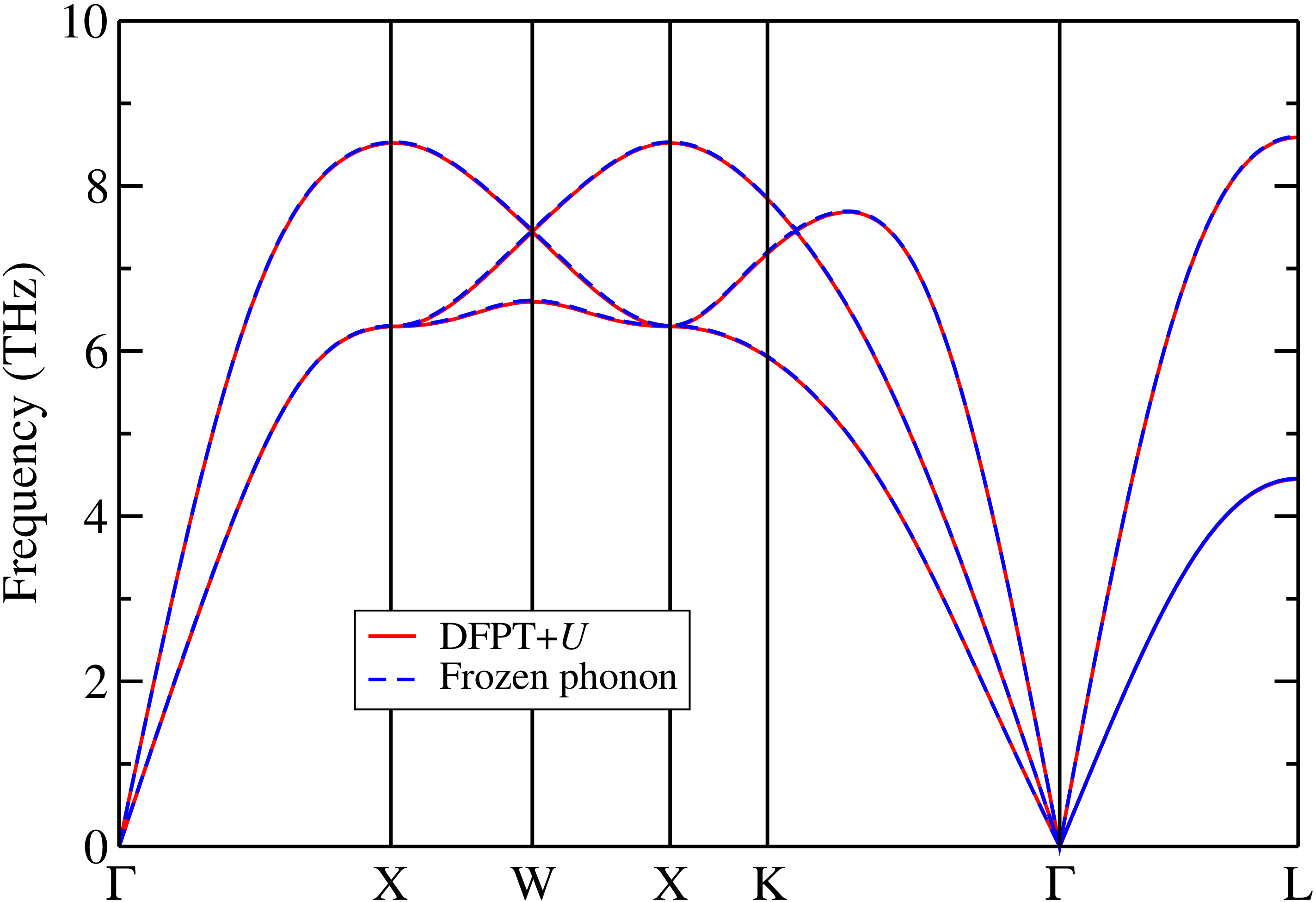}}
  \caption{Comparison of the phonon dispersion for bulk metallic Ni computed with DFPT+$U$ and the with frozen-phonon approach based on DFT+$U$ forces in supercells. Calculations were performed using a US PP (a), and a PAW PP (b).}
  \label{figSM:Ni_disp}
\end{center}
\end{figure*}

{\bf Discussion.} 
The aim of this section is to benchmark the DFPT+$U$ implementation for metallic systems. Again, we test both US and PAW PPs. The benchmark is realized by contrasting the phonon dispersions obtained from DFPT+$U$ against those from frozen-phonon calculations. 
Bulk \textit{fcc} Ni was selected as a test case.
Although for this system GGA already gives a phonon dispersion in 
very good
agreement with  inelastic neutron scattering data~[8], the finite effect that the Hubbard correction has on the 
phonon frequencies will 
represent a significant test for DFPT+$U$ on metallic ground states. 
Fig.~\ref{figSM:Ni_disp} shows a comparison of the phonon dispersion of bulk Ni as computed from DFPT+$U$ and  the frozen-phonon approach. The two 
computational schemes produce   
phonon dispersions
superimposed to each other 
providing a solid demonstration of the correctness of our DFPT+$U$ implementation also for metals (i.e., in presence of fractionally occupied KS states). 

\vskip 0.2cm

{\bf Technical details.} The  calculations 
were performed using the 
experimental lattice parameter of  3.499~\AA~[9]. The exchange-correlation energy was modeled using a PBEsol functional~[10], while both the US and PAW PPs were taken from the PSlibrary~0.3.1~[11]. 
Kinetic-energy cutoffs of 80~Ry and 960~Ry were used for KS wave functions and charge-density, respectively. We used the Marzari-Vanderbilt smearing technique~[12] with a broadening parameter of 0.02~Ry. DFPT+$U$ calculations were performed on uniform $\Gamma$-centered $\mathbf{k}$ and $\mathbf{q}$ meshes of $10 \times 10 \times 10$ and $5 \times 5 \times 5$ points, respectively. Consistently with these settings, frozen-phonon calculations employed a $5 \times 5 \times 5$ supercell  
and a  $2 \times 2 \times 2$   $\mathbf{k}$ point mesh. 
Frozen-phonon pre- and post-processing analysis were performed using the Phonopy code~[3]. 
We  used a test Hubbard $U=2$~eV 
for Ni $3d$ states both for US and PAW PPs.

\vskip 0.3cm

\clearpage

\newpage

\noindent
[1] Ultrasoft pseudopotentials from the GBRV library 1.2 in K. Garrity, J. Bennett, K. Rabe, and D. Vanderbilt, Comput. Mater. Sci. {\bf 81}, 446 (2014) : \texttt{co\_pbesol\_v1.2.uspp.F.UPF} and \texttt{o\_pbesol\_v1.2.uspp.F.UPF}.

\noindent
[2] Projector augmented wave pseudopotentials from the PSlibrary 0.3.1 in A. Dal Corso, Comput. Mater. Sci.95, {\bf 337} (2014): \texttt{Co.pbesol-spn-kjpaw psl.0.3.1.UPF}, \texttt{O.pbesol-n-kjpaw psl.0.1.UPF}, and \texttt{Li.pbesol-s-kjpaw psl.0.2.1.UPF}.

\noindent
[3] A. Togo and I. Tanaka, Scr. Mater. {\bf 108}, 1 (2015).

\noindent
[4] P. Giannozzi, S. Baroni, N. Bonini, M. Calandra, R. Car, C. Cavazzoni, D. Ceresoli, G. Chiarotti, M. Cococcioni, I.~Dabo, A. Dal Corso, S. De Gironcoli, S. Fabris, G. Fratesi, R. Gebauer, U. Gerstmann, C. Gougoussis, A. Kokalj, M. Lazzeri, L. Martin-Samos, N. Marzari, F. Mauri, R. Mazzarello, S. Paolini, A. Pasquarello, L. Paulatto, C. Sbraccia, S. Scandolo, G. Sclauzero, A. Seitsonen, A. Smogunov, P. Umari, and R. Wentzcovitch, J. Phys.: Condens. Matter {\bf 21}, 395502 (2009).

\noindent
[5] P. Giannozzi, O. Andreussi, T. Brumme, O. Bunau, M. Buongiorno Nardelli, M. Calandra, R. Car, C. Cavazzoni, D. Ceresoli, M. Cococcioni, N. Colonna, I. Carnimeo, A. Dal Corso, S. de Gironcoli, P. Delugas, R. A. DiStasio~Jr., A. Ferretti, A. Floris, G. Fratesi, G. Fugallo, R. Gebauer, U. Gerstmann, F. Giustino, T. Gorni, J. Jia, M. Kawamura, H.-Y. Ko, A. Kokalj, E. K\"{u}\c{c}\"{u}kbenli, M. Lazzeri, M. Marsili, N. Marzari, F. Mauri, N. L. Nguyen, H.-V. Nguyen, A. Otero-de-la Rosa, L. Paulatto, S. Ponc\'e, D. Rocca, R. Sabatini, B. Santra, M. Schlipf, A. Seitsonen, A. Smogunov, I. Timrov, T. Thonhauser, P. Umari, N. Vast, and S. Baroni, J. Phys.: Condens. Matter {\bf 29}, 465901 (2017).

\noindent
[6] S. Baroni, S. de Gironcoli, A. Dal Corso, and P. Giannozzi, Rev. Mod. Phys. {\bf 73}, 515 (2001).

\noindent
[7] P. Giannozzi, S. de Gironcoli, P. Pavone, and S. Baroni, Phys. Rev. B {\bf 43}, 7231 (1991).

\noindent
[8] A. Dal Corso and S. de Gironcoli, Phys. Rev. B {\bf 62}, 273 (2000).

\noindent
[9] W. Davey, Phys. Rev. {\bf 25}, 753 (1925).

\noindent
[10] J. P. Perdew, A. Ruzsinszky, G. I. Csonka, O. A. Vydrov, G. E. Scuseria, L. A. Constantin, X. Zhou, and K.~Burke, Phys. Rev. Lett. {\bf 100}, 136406 (2008).

\noindent
[11] Ultrasoft and projector augmented wave pseudopotentials for Ni from the PSlibrary 0.3.1 in A. Dal Corso, Comput. Mater. Sci. {\bf 95}, 337 (2014): \texttt{Ni.pbesol-spn-rrkjus psl.0.3.1.UPF} and \texttt{Ni.pbesol-spn-kjpaw psl.0.3.1.UPF}.

\noindent
[12] N. Marzari, D. Vanderbilt, A. De Vita, and M. C. Payne, Phys. Rev. Lett. {\bf 82}, 3296 (1999).

\end{document}